\documentclass[preprint,10pt]{elsarticle}
\pdfoutput=1
\pdfcatalog{/PageLayout /SinglePage}



\makeatletter
\def\ps@pprintTitle{%
   \let\@oddhead\@empty
   \let\@evenhead\@empty
   \def\@oddfoot{\reset@font\hfil\thepage\hfil}
   \let\@evenfoot\@oddfoot
}
\makeatother

\usepackage{titlesec}
\usepackage{xcolor}
\usepackage{enumitem}
\usepackage{hyperref}
\hypersetup{colorlinks=true,urlcolor=black,pdfpagemode=UseNone,pdfstartview=Fit}
\usepackage[T1]{fontenc}\usepackage{lmodern}

\usepackage{mathtools,amsfonts,amssymb}
\usepackage{old-arrows}
\usepackage{caption}
\usepackage{subcaption}
\usepackage{booktabs}
\usepackage{pifont}
\usepackage{placeins}

\definecolor{green}{rgb}{0,0.5,0}

\usepackage{tikz}
\usetikzlibrary{calc}
\usepackage{tkz-euclide}
\usepackage{calc}
\usepackage{tikz,pgfplots}
\pgfplotsset{compat=1.12}
\usetikzlibrary{arrows.meta,3d,matrix,positioning,decorations.markings,math}
\makeatletter
\tikzoption{canvas is xy plane at z}[]{%
  \def\tikz@plane@origin{\pgfpointxyz{0}{0}{#1}}%
  \def\tikz@plane@x{\pgfpointxyz{1}{0}{#1}}%
  \def\tikz@plane@y{\pgfpointxyz{0}{1}{#1}}%
  \tikz@canvas@is@plane
}
\makeatother

\makeatletter
\tikzoption{canvas is plane}[]{\@setOxy#1}
\def\@setOxy O(#1,#2,#3)x(#4,#5,#6)y(#7,#8,#9)%
  {\def\tikz@plane@origin{\pgfpointxyz{#1}{#2}{#3}}%
   \def\tikz@plane@x{\pgfpointxyz{#4}{#5}{#6}}%
   \def\tikz@plane@y{\pgfpointxyz{#7}{#8}{#9}}%
   \tikz@canvas@is@plane
  }
\makeatother

\usepackage{xfrac}

\newcommand{\pz}{\phantom{0}}

\newcommand{\vpad}{\vphantom{\bigg(}}
\newcommand{\ntriangles}{n_t}

\newcommand{\nbasis}{n_b}
\newcommand{\srcidx}{j}
\newcommand{\testidx}{i}
\DeclareMathOperator{\real}{Re}

\newcommand{\aee}{a_{\mathcal{E},\mathcal{E}}} 
\newcommand{\amm}{a_{\mathcal{M},\mathcal{M}}} 
\newcommand{\ame}{a_{\mathcal{M},\mathcal{E}}} 
\newcommand{\aem}{a_{\mathcal{E},\mathcal{M}}} 
\newcommand{\be}{b_\mathcal{E}} 

\newcommand{\lint}{L^\text{int}}
\newcommand{\lext}{L^\text{ext}}
\newcommand{\aext}{a^\text{ext}}
\newcommand{\bext}{b^\text{ext}}
\newcommand{\aint}{a^\text{int}}
\newcommand{\bint}{b^\text{int}}
\newcommand{\cint}{c^\text{int}}

\newcommand{\aslot}{a^\text{slot}}
\newcommand{\bslot}{b^\text{slot}}
\newcommand{\lslot}{L}
\newcommand{\xslot}{x_w^\text{int}}
\newcommand{\solvec}{\boldsymbol{\mathcal{J}}^h}

\newcommand{\WidestEntryD}{$\mathbf{D}$}%
\newcommand{\SetToWidestD}[1]{\makebox[\widthof{\WidestEntryD}]{$#1$}}%
\newcommand{\WidestEntryExt}{$^-$}%
\newcommand{\SetToWidestExt}[1]{\makebox[\widthof{\WidestEntryExt}][l]{$#1$}}%
\newcommand{\WidestEntryInt}{$^+$}%
\newcommand{\SetToWidestInt}[1]{\makebox[\widthof{\WidestEntryInt}][l]{$#1$}}%

\definecolor{orange}{rgb}{1,0.5,0}
\definecolor{green}{rgb}{0,0.5,0}
\definecolor{purple}{rgb}{0.5,0,0.5}

\newcommand{\rereading}[1]{#1}

\usepackage[margin=1in]{geometry}
\biboptions{sort&compress}
\begin{document}

\begin{frontmatter}
\title{Code-Verification Techniques for an Arbitrary-Depth Electromagnetic Slot Model}

\author[freno]{Brian A.~Freno}
\ead{bafreno@sandia.gov}
\author[freno]{Neil R.~Matula}
\author[freno]{Robert A.~Pfeiffer}
\author[freno]{Vinh Q.~Dang}
\address[freno]{Sandia National Laboratories, Albuquerque, NM 87185}

\begin{abstract}
Electromagnetic slot models are employed to efficiently simulate electromagnetic penetration through openings in an otherwise closed electromagnetic scatterer.  Such models, which incorporate varying assumptions about the geometry of the openings, are typically coupled with electromagnetic surface integral equations that model electromagnetic scattering.  In this paper, we introduce novel code-verification approaches and build upon our previously developed methodologies to assess the correctness of the numerical implementation of an arbitrary-depth slot model.  Through these approaches, we measure the convergence rates of the different interacting sources of numerical error and demonstrate the impact of various factors on these rates for several cases.

\end{abstract}


\end{frontmatter}

\section{Introduction}

A frequently encountered problem in computational electromagnetics is the presence of imperfectly sealed gaps in electromagnetic shielding~\cite{butler_1978}.  Through these openings, the exterior electromagnetic field interacts with the interior, preventing the interior from being fully shielded.  To efficiently simulate electromagnetic penetration, slot models are typically coupled with electromagnetic surface integral equations, such as the electric-, magnetic-, and combined-field equations.  \rereading{The numerical evaluation of surface integral equations incurs a significantly lower computational cost than that of volume-based methods, while better accommodating more general geometries. The slots considered} are typically assumed to be rectangular prisms with assumptions on the relative size of one or more dimensions.  The width of the slot is typically small compared to the overall size of the scatterer.  Therefore, to avoid the computational burden of resolving the small length scales in the vicinity and interior of the slot, the effect of the slot may be modeled by carefully chosen source currents affixed to the scattering surface~\cite{warne_1988}.  Slot model development and validation  remain active research topics~\cite{schelkunoff_1952,cerri_1992,jedlicka_1995,robinson_1998,araneo_2008,hill_2009,pozar_2011,campione_2020,warne_2020,illescas_2023}.

For computational physics codes, code verification is critical for evaluating whether the numerical algorithms have been correctly implemented~\cite{roache_1998,knupp_2022,oberkampf_2010}.  The discretization of differential and integral operators introduces a discretization error in the numerical solution. The numerical implementation of these operators can be verified by measuring how quickly the error decreases with discretization refinement and comparing with the expected rate for numerous test cases.  To compute the error, the method of manufactured solutions~\cite{roache_2001} is commonly used to create problems of arbitrary complexity with known solutions.  
Code-verification examples have been published for computational mechanics and heat transfer~\cite{roy_2004,bond_2007,veluri_2010,chamberland_2010,amar_2011,oliver_2012,etienne_2012,veeraragavan_2016,eca_2016,hennink_2021,freno_2021,freno_ablation_2022,bukac_2023,bukreev_2023,fumagalli_2024} and computational electromagnetics and plasma sciences~\cite{mcclarren_2008,riva_2017,amormartin_2021,tranquilli_2022,rueda_2023,rudi_2024,issan_2024}.
Code-verification approaches for electromagnetic surface integral equations have been developed for the electric-field integral equation~\cite{marchand_2013,marchand_2014,freno_em_mms_2020,freno_em_mms_quad_2021,freno_efie_slot_2024}, magnetic-field integral equation~\cite{freno_mfie_2022}, and combined-field integral equation~\cite{freno_cfie_2023}.  

In this work, we introduce novel code-verification approaches and build upon our previously developed methodologies for a slot model that can accommodate an arbitrary depth, as described in~\cite{warne_2020}.  This model differs from the \textit{thick} (i.e., small-depth) slot model considered in~\cite{freno_efie_slot_2024} that is described in~\cite{warne_1990,warne_1992,warne_1995,johnson_2002}.  Unlike the thick slot model, which assumes the magnetic currents along the two aperture wires are equal and opposite due to the small depth, the model considered in this paper permits these currents to differ, and therefore assigns a separate set of degrees of freedom to the currents on each wire.

As described in~\cite{freno_em_mms_2020}, codes for solving electromagnetic surface integral equations incur numerical error from multiple sources, including faceted approximations of curved surfaces, finite-dimensional solution spaces, and approximate integration.  In addition to these sources, the arbitrary-depth slot model studied here introduces further error from truncated series approximations.  In this paper, we provide code-verification techniques to either verify the convergence rates of each of these error contributions, or eliminate their effect.

The structure of this paper is outlined as follows.  We provide the equations for the surface of the electromagnetic scatterer and the arbitrary-depth slot model in Section~\ref{sec:equations}, and we describe how they are discretized in Section~\ref{sec:discretization}. We discuss our code-verification approaches for these equations and their expected convergence rates in Section~\ref{sec:mms}.  For multiple examples, we illustrate the efficacy of these methods and impact of the series truncation on the convergence rate in Section~\ref{sec:results}.  We summarize our work in Section~\ref{sec:conclusions}.
\begin{figure}
\centering
\input{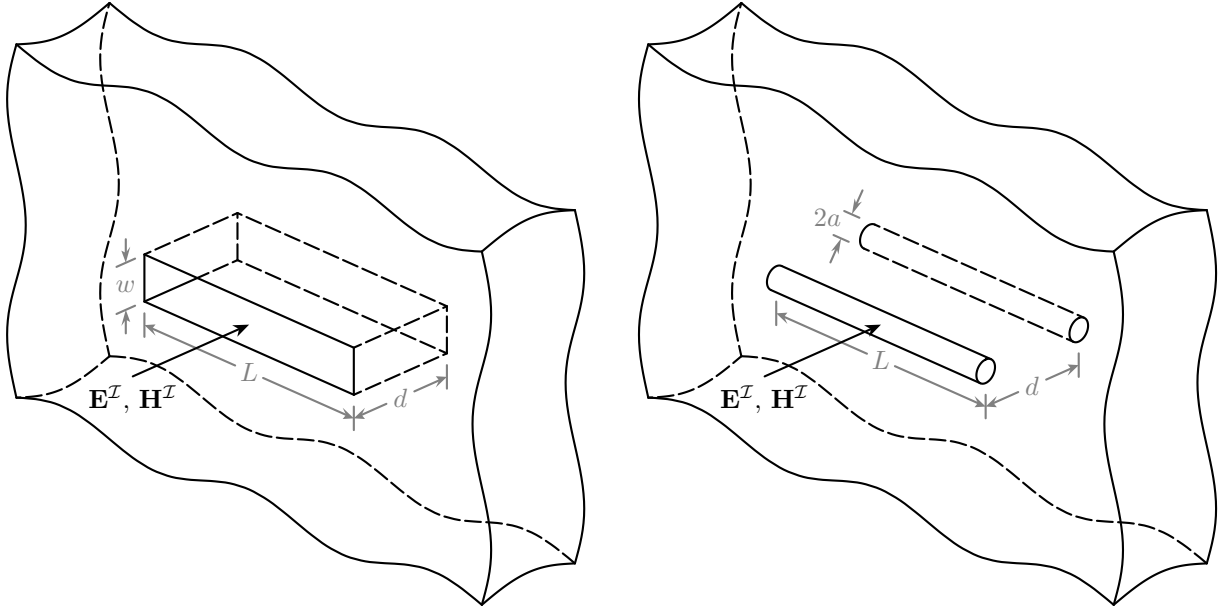}
\caption{
The exterior of the electromagnetic scatterer is connected to the interior of the cavity by a slot (left), which is modeled by a pair of wires positioned along the slot openings (right)~\cite{freno_efie_slot_2024}.}
\vskip-\dp\strutbox
\label{fig:slot}
\end{figure}

\newcommand{\modeidx}{p}
\newcommand{\modeidxp}{q}
\newcommand{\modemaxp}{Q}

\section{Governing Equations}
\label{sec:equations}

We focus on a narrow, rectangularly prismatic slot that otherwise prevents an electromagnetic scatterer from fully enclosing a cavity, as shown in Figure~\ref{fig:slot}.  The width $w$ of the slot is assumed to be much smaller than its length $L$.  However, unlike the slot considered in~\cite{freno_efie_slot_2024}, the depth $d$ of the slot is of an arbitrary extent.  Aside from the slot, the scatterer exterior and the cavity interior are represented as distinct closed surfaces using the electric-field integral equation (EFIE).  At its openings, the slot is modeled by thin wires that conduct magnetic current.  The surfaces of the scatterer exterior and cavity interior interact with their respective wires rather than interacting directly with one another, and the wires interact with each other through a waveguide model.

The electric current on the surfaces of the scatterer exterior and cavity interior is modeled using the EFIE for a moderately resistive conductor~\cite{freno_efie_slot_2024}.  
The problem can be expressed in its variational form, where we seek the electric surface current density $\mathbf{J}\in\mathbb{V}$ and wire magnetic current $\mathbf{I}_m=I_m(s)\mathbf{s}\in\mathbb{V}^m$ that satisfy
\begin{align}
\aee(\mathbf{J},\mathbf{v}) + \aem(\mathbf{I}_m,\mathbf{v}) = \be\bigl(\mathbf{E}^\mathcal{I}, \mathbf{v}\bigr)
\label{eq:efie_sesquilinear}
\end{align}
for every $\mathbf{v}\in\mathbb{V}$, where $\mathbb{V}$ represents the space of vector fields that are tangent to the surface $S'=S$.  The prime distinguishes the domains for the source and test integrals.
$\mathbb{V}^m$ represents the space of vector fields that are located on and tangent to the wire and disappear at the wire endpoints $s=0$ and $s=\lslot$, where $s\in[0,\,\lslot]$ denotes the position along the wire, and $\mathbf{s}$ indicates the wire direction.
The operators in~\eqref{eq:efie_sesquilinear} are defined by
\begin{align}
\aee(\mathbf{u},\mathbf{v}) ={}& 
-\frac{j}{\omega\epsilon} \int_S \nabla\cdot\bar{\mathbf{v}}(\mathbf{x})\int_{S'} \nabla'\cdot\mathbf{u}(\mathbf{x}')G(\mathbf{x},\mathbf{x}')dS' dS
{}+{}
j\omega\mu \int_S \bar{\mathbf{v}}(\mathbf{x})\cdot\int_{S'} \mathbf{u}(\mathbf{x}')G(\mathbf{x},\mathbf{x}')dS'dS 
\nonumber \\ &+ Z_s\int_S \bar{\mathbf{v}}(\mathbf{x})\cdot \mathbf{u}(\mathbf{x}) dS, \label{eq:aee}
\\
\aem(\mathbf{u},\mathbf{v}) ={}& -\frac{1}{4}\int_0^\lslot \bar{\mathbf{v}}(\mathbf{x})\cdot \bigl[\mathbf{n}(\mathbf{x})\times \mathbf{u}(s)\bigr] ds + \frac{1}{4\pi}\int_S \bar{\mathbf{v}}(\mathbf{x})\cdot  \int_0^\lslot \mathbf{u}(s')\times\int_0^{2\pi} \nabla' G(\mathbf{x},\mathbf{x}')d\phi' ds' dS,
\label{eq:aem}
\\
\be(\mathbf{u},\mathbf{v})  ={}& \int_S \bar{\mathbf{v}}(\mathbf{x})\cdot\mathbf{u}(\mathbf{x})  dS. \label{eq:be}
\end{align}
In~\eqref{eq:aee}--\eqref{eq:be}, the overbar indicates complex conjugation; 
\begin{align}
G(\mathbf{x},\mathbf{x}') = \frac{e^{-jkR}}{4\pi R}
\label{eq:G}
\end{align}
is the Green's function, where $R=\|\mathbf{x}-\mathbf{x}'\|_2$ is the distance between the test and source points and $k=\omega\sqrt{\mu\epsilon}$ is the wavenumber; $\omega$ is the angular frequency; $\mu$ and $\epsilon$ are the permeability and permittivity of the medium that surrounds the surface; $Z_s$ is the resistive surface impedance of the conductor; and $\mathbf{n}$ is the unit vector that is normal to the surface of the conductor and points away from the conductor.

The slot is modeled as a rectangular waveguide with an electrically small width~\cite{warne_2020}, and the details of its derivation are included in Appendix~\ref{app:slot_model}.
The problem can be expressed in its variational form, where we seek the wire magnetic current $\mathbf{I}_m\in\mathbb{V}^m$ and the electric surface current density $\mathbf{J}\in\mathbb{V}$ that satisfy
\begin{align}
\ame(\mathbf{J},\mathbf{v}^m) + \amm(\mathbf{I}_m,\mathbf{v}^m) = 0
\label{eq:slot_sesquilinear}
\end{align}
for every $\mathbf{v}^m\in\mathbb{V}^m$.
The operators are defined by
\begin{align}
\ame(\mathbf{u},\mathbf{v}) ={}& \int_0^\lslot \bar{\mathbf{v}}(s)\cdot\bigl[\mathbf{u}(\mathbf{x})\times\mathbf{n}(\mathbf{x})\bigr] ds, \nonumber
\\
\amm^{\sim}    (\mathbf{u},\mathbf{v})  ={}& \frac{j\omega\epsilon}{2 w L \bigl(k^2 -\beta_x^2\bigr)}
\sum_{\modeidx=1}^\infty
\beta_{y_\modeidx}F^p(\bar{\mathbf{v}}) F^p(\mathbf{u})
\Bigl(-\tan(\beta_{y_\modeidx} d/2) + \cot(\beta_{y_\modeidx} d/2)\Bigr), \label{eq:ammsim}
\\
\amm^{\not\sim}(\mathbf{u},\mathbf{v})  ={}& \frac{j\omega\epsilon}{2 w L \bigl(k^2 -\beta_x^2\bigr)}
\sum_{\modeidx=1}^\infty
\beta_{y_\modeidx}F^p(\bar{\mathbf{v}}) F^p(\mathbf{u})
\Bigl(+\tan(\beta_{y_\modeidx} d/2) + \cot(\beta_{y_\modeidx} d/2)\Bigr), \label{eq:ammdissim}
\end{align}
where
\begin{align*}
F^p(\mathbf{u}) = \int_0^\lslot (\mathbf{u}(s)\cdot\mathbf{s})\sin \biggl(\frac{\modeidx\pi s}{\lslot}\biggr)ds.
\end{align*}
For $\amm$, the superscript ($\sim$) indicates $\mathbf{u}$ and $\mathbf{v}$ are located at the same opening, whereas the superscript ($\not\sim$) indicates opposite openings.
In~\eqref{eq:ammsim} and~\eqref{eq:ammdissim}, 
\begin{align}
\beta_x = \sqrt{\frac{2 j Z_s \omega \epsilon}{w}}, \qquad 
\beta_{y_\modeidx} = \sqrt{k^2 -\beta_x^2 - \beta_{z_\modeidx}^2}, \qquad 
\beta_{z_\modeidx} =  \frac{\modeidx\pi}{\lslot}
\label{eq:prop_const}
\end{align}
are the propagation constants, and $k$ and $\epsilon$ are the potentially complex wavenumber and permittivity associated with the medium that occupies the slot interior.

\section{Discretization} 
\label{sec:discretization}

We solve \eqref{eq:efie_sesquilinear} and \eqref{eq:slot_sesquilinear} by discretizing the surfaces with triangular elements and the wires with one-dimensional bar elements.  We approximate $\mathbf{J}$ with $\mathbf{J}_h$ using the Rao--Wilton--Glisson (RWG) basis functions $\boldsymbol{\Lambda}_{\srcidx}(\mathbf{x})$~\cite{rao_1982} and $\mathbf{I}_m$ with $\mathbf{I}_h$ using the one-dimensional analog $\boldsymbol{\Lambda}_{\srcidx}^m(s)$~\cite{freno_efie_slot_2024}:
\begin{align}
\mathbf{J}_h(\mathbf{x}) = \sum_{\srcidx=1}^{\nbasis} J_{\srcidx} \boldsymbol{\Lambda}_{\srcidx}(\mathbf{x}),
\qquad
\mathbf{I}_h(s) = \sum_{\srcidx=1}^{\nbasis^m} I_{\srcidx} \boldsymbol{\Lambda}_{\srcidx}^m(s),
\label{eq:sol_h}
\end{align}%
where $\nbasis$ and $\nbasis^m$ are the numbers of RWG and one-dimensional basis functions. For the RWG basis functions, we measure the solution at the edge midpoints; for the one-dimensional basis functions, we measure the solution at the nodes.

Letting $\mathbb{V}_h$ and $\mathbb{V}_h^m$ denote the span of RWG and one-dimensional basis functions and inserting~\eqref{eq:sol_h} into~\eqref{eq:efie_sesquilinear} and~\eqref{eq:slot_sesquilinear}, we seek $\mathbf{J}_h\in\mathbb{V}_h$ and $\mathbf{I}_h\in\mathbb{V}_h^m$ that satisfy
\begin{align}
\aee(\mathbf{J}_h,\boldsymbol{\Lambda}_{\testidx}) + 
\aem(\mathbf{I}_h,\boldsymbol{\Lambda}_{\testidx}) = \be\bigl(\mathbf{E}^\mathcal{I}, \boldsymbol{\Lambda}_{\testidx}\bigr)
\label{eq:proj_disc_efie}
\end{align}
for $i=1,\hdots,\nbasis$, and 
\begin{align}
\ame(\mathbf{J}_h,\boldsymbol{\Lambda}_{\testidx}^m) + \amm(\mathbf{I}_h,\boldsymbol{\Lambda}_{\testidx}^m) = 0
\label{eq:proj_disc_slot}
\end{align}
for $i=1,\hdots,\nbasis^m$.

We evaluate~\eqref{eq:proj_disc_efie} on the surfaces of the scatterer exterior ($-$) and cavity interior ($+$) to solve for the $\nbasis=\nbasis^-+\nbasis^+$ unknowns for $\mathbf{J}_h$.  Likewise, we evaluate~\eqref{eq:proj_disc_slot} for the corresponding wires.  
In this work, we model the slot as having an arbitrary depth, such that the two wires are modeled with separate unknowns but with the same number of unknowns per wire; consequently, there are $\nbasis^m={\nbasis^m}^- + {\nbasis^m}^+$ unknowns for $\mathbf{I}_h$, and ${\nbasis^m}^-={\nbasis^m}^+$.  This modeling paradigm differs from that described in~\cite{freno_efie_slot_2024}, where both wires are modeled with the same unknowns.

In matrix--vector form, we write~\eqref{eq:proj_disc_efie} and~\eqref{eq:proj_disc_slot} as
\begin{align}
\mathbf{Z}\solvec = \mathbf{V}.
\label{eq:system}
\end{align}
We write impedance matrix as
\begin{align*}
\mathbf{Z} = \left[\begin{array}{@{} c @{} l r @{} c @{} l c @{} l r @{} c @{} l @{}}
\mathbf{A}&^- &  &\mathbf{0} &   & \mathbf{B} &^-            &  &\mathbf{0} &              \\ 
\mathbf{0}&   &  &\mathbf{A} &^+ & \mathbf{0} &              &  &\mathbf{B} &^+            \\ 
\mathbf{C}&^- &  &\mathbf{0} &   & \mathbf{D} &_\sim^-       &  &\mathbf{D} &_{\not\sim}^- \\ 
\mathbf{0}&   &  &\mathbf{C} &^+ & \mathbf{D} &_{\not\sim}^+ &  &\mathbf{D} &_\sim^+ 
\end{array}\right]\in\mathbb{C}^{(\nbasis+\nbasis^m)\times(\nbasis+\nbasis^m)},
\end{align*}
where  
\begin{alignat*}{9}
A_{\testidx,\srcidx}&{}={}&\pz\aee(\boldsymbol{\Lambda}_{\srcidx}^{\phantom{m}},\boldsymbol{\Lambda}_{\testidx}^{\phantom{m}}),  &&\qquad
\mathbf{A}^-\in\mathbb{C}&^{\nbasis^-} &&^{\times\nbasis^-}&&, \qquad &
\mathbf{A}^+\in\mathbb{C}&^{\nbasis^+} &&^{\times\nbasis^+}&&,
\\
B_{\testidx,\srcidx}&{}={}&\pz\aem(\boldsymbol{\Lambda}_{\srcidx}^m ,\boldsymbol{\Lambda}_{\testidx}^{\phantom{m}}), &&\qquad
\mathbf{B}^-\in\mathbb{C}&^{\nbasis^-} &&^{\times{\nbasis^m}^-}&&,  \qquad &
\mathbf{B}^+\in\mathbb{C}&^{\nbasis^+} &&^{\times{\nbasis^m}^+}&&,
\\
C_{\testidx,\srcidx}&{}={}&\pz\ame(\boldsymbol{\Lambda}_{\srcidx}^{\phantom{m}},\boldsymbol{\Lambda}_{\testidx}^m ), &&\qquad
\mathbf{C}^-\in\mathbb{R}&^{{\nbasis^m}^-} &&^{\times\nbasis^-}&&,  \qquad &
\mathbf{C}^+\in\mathbb{R}&^{{\nbasis^m}^+} &&^{\times\nbasis^+}&&,
\\
{D_\sim}_{\testidx,\srcidx} &{}={}&  \amm^\sim(\boldsymbol{\Lambda}_{\srcidx}^m ,\boldsymbol{\Lambda}_{\testidx}^m ),  &&\qquad
\mathbf{D}_\sim^-      \in\mathbb{C}&^{{\nbasis^m}^-} &&^{\times{\nbasis^m}^-}&&,\qquad &
\mathbf{D}_\sim^+      \in\mathbb{C}&^{{\nbasis^m}^+} &&^{\times{\nbasis^m}^+}&&,
\\
{D_{\not\sim}}_{\testidx,\srcidx} &{}={}&  \amm^{\not\sim}(\boldsymbol{\Lambda}_{\srcidx}^m ,\boldsymbol{\Lambda}_{\testidx}^m ),  &&\qquad
\mathbf{D}_{\not\sim}^+\in\mathbb{C}&^{{\nbasis^m}^+} &&^{\times{\nbasis^m}^-}&&,\qquad &
\mathbf{D}_{\not\sim}^-\in\mathbb{C}&^{{\nbasis^m}^-} &&^{\times{\nbasis^m}^+}&&.
\end{alignat*}
More succinctly, we write $\mathbf{Z}$ as
\begin{align}
\mathbf{Z} = \left[\begin{matrix}
\mathbf{A} & \mathbf{B} \\ 
\mathbf{C} & \mathbf{D} \end{matrix}\right],\label{eq:Z}
\end{align}
where 
\begin{alignat*}{7}
\SetToWidestD{\mathbf{A}}&{}= \left[\begin{array}{@{} c @{} l c @{} l @{}}
\SetToWidestD{\mathbf{A}}&^- & \SetToWidestD{\mathbf{0}} \\ 
\SetToWidestD{\mathbf{0}}&   & \SetToWidestD{\mathbf{A}} &^+ \end{array}\right] \in\mathbb{C}&&^{\nbasis}&&^{\times\nbasis},
\qquad
&
\SetToWidestD{\mathbf{B}}&{}= \left[\begin{array}{@{} c @{} l c @{} l @{}}
\SetToWidestD{\mathbf{B}}&^- & \SetToWidestD{\mathbf{0}} \\ 
\SetToWidestD{\mathbf{0}}&   & \SetToWidestD{\mathbf{B}} &^+ \end{array}\right]\in\mathbb{C}&&^{\nbasis}&&^{\times\nbasis^m},
\\
\SetToWidestD{\mathbf{C}}&{}= \left[\begin{array}{@{} c @{} l c @{} l @{}}
\SetToWidestD{\mathbf{C}}&^- & \SetToWidestD{\mathbf{0}} \\ 
\SetToWidestD{\mathbf{0}}&   & \SetToWidestD{\mathbf{C}} &^+ \end{array}\right]\in\mathbb{R}&&^{\nbasis^m}&&^{\times\nbasis},
\qquad
&
\SetToWidestD{\mathbf{D}}&{}= \left[\begin{array}{@{} c @{} l c @{} l @{}}
\SetToWidestD{\mathbf{D}}&\SetToWidestExt{_\sim^-} & \SetToWidestD{\mathbf{D}} &\SetToWidestInt{_{\not\sim}^-} \\ 
\SetToWidestD{\mathbf{D}}&\SetToWidestExt{_{\not\sim}^+} & \SetToWidestD{\mathbf{D}} &\SetToWidestInt{_\sim^+} \end{array}\right]\in\mathbb{C}&&^{\nbasis^m}&&^{\times\nbasis^m}.
\end{alignat*}
We write the solution vector containing the coefficients in~\eqref{eq:sol_h} as
\begin{align*}
\solvec = \left\{\begin{array}{@{} r @{} l @{}}
\mathbf{J}&^h\strut^- \\ 
\mathbf{J}&^h\strut^+ \\ 
\mathbf{I}&^h\strut^- \\ 
\mathbf{I}&^h\strut^+
\end{array}\right\}\in\mathbb{C}^{\nbasis+\nbasis^m},
\end{align*}
where
\begin{alignat*}{9}
J&_{\srcidx}^h &&{}={} &J_{\srcidx},  &&\qquad \mathbf{J}&^h\strut^-&&{}\in\mathbb{C}^{\nbasis^-}&&, \qquad & {\mathbf{J}^h}^+&{}\in\mathbb{C}^{\nbasis^+},
\\
I&_{\srcidx}^h &&{}={} &I_{\srcidx},  &&\qquad \mathbf{I}&^h\strut^-&&{}\in\mathbb{C}^{{\nbasis^m}^-}&&, \qquad & {\mathbf{I}^h}^+&{}\in\mathbb{C}^{{\nbasis^m}^+}.
\end{alignat*}
More succinctly, we write $\solvec$ as
\begin{align*}
\solvec = \left\{\begin{array}{@{} r @{} l @{}}
\mathbf{J}&^h \\ 
\mathbf{I}&^h\end{array}\right\},
\end{align*}
where
\begin{align*}
\mathbf{J}^h = \left\{\begin{array}{@{} r @{} l @{}}
\mathbf{J}&^h\strut^- \\ 
\mathbf{J}&^h\strut^+\end{array}\right\}\in\mathbb{C}^{\nbasis},
\qquad
\mathbf{I}^h = \left\{\begin{array}{@{} r @{} l @{}}
\mathbf{I}&^h\strut^- \\ 
\mathbf{I}&^h\strut^+\end{array}\right\}\in\mathbb{C}^{\nbasis^m}
.
\end{align*}
Finally, we write the excitation vector as
\begin{align*}
\mathbf{V} = \left\{\begin{array}{@{} c @{} l @{}}
\mathbf{V}& ^\mathcal{E}\strut^- \\ 
\mathbf{V}& ^\mathcal{E}\strut^+ \\ 
\mathbf{0}& \end{array}\right\}\in\mathbb{C}^{\nbasis+\nbasis^m},
\end{align*}
where
\begin{alignat*}{7}
V^\mathcal{E}_{\testidx} &{}= \be\bigl(\mathbf{E}^\mathcal{I} , \boldsymbol{\Lambda}_{\testidx}\bigr),  \qquad &
{\mathbf{V}^\mathcal{E}}^-&{}\in\mathbb{C}^{\nbasis^-}, \qquad &
{\mathbf{V}^\mathcal{E}}^+&{}\in\mathbb{C}^{\nbasis^+}.
\end{alignat*}
%

\section{Manufactured Solutions} 
\label{sec:mms}

The residual functionals for~\eqref{eq:efie_sesquilinear} and~\eqref{eq:slot_sesquilinear} are
\begin{alignat}{7}
r_{\mathcal{E}_{\testidx}}(\mathbf{u},\mathbf{v}) &{}={}& \aee(\mathbf{u},\boldsymbol{\Lambda}_{\testidx}^{\phantom{m}}) &{}+{}& \aem(\mathbf{v},\boldsymbol{\Lambda}_{\testidx}^{\phantom{m}}) &{}-{}&\be\bigl(\mathbf{E}^\mathcal{I}, \boldsymbol{\Lambda}_{\testidx}\bigr),
\label{eq:res_func_efie}
\\
r_{\mathcal{M}_{\testidx}}(\mathbf{u},\mathbf{v}) &{}={}& \ame(\mathbf{u},\boldsymbol{\Lambda}_{\testidx}^m) &{}+{}& \amm(\mathbf{v},\boldsymbol{\Lambda}_{\testidx}^m)&.
\label{eq:res_func_slot}
\end{alignat}
Using~\eqref{eq:res_func_efie} and~\eqref{eq:res_func_slot}, the variational forms of~\eqref{eq:efie_sesquilinear} and~\eqref{eq:slot_sesquilinear} are
\begin{alignat}{7}
r_{\mathcal{E}_{\testidx}}(\mathbf{J},\mathbf{I}_m) &{}={}& \aee(\mathbf{J},\boldsymbol{\Lambda}_{\testidx}^{\phantom{m}}) &{}+{}& \aem(\mathbf{I}_m,\boldsymbol{\Lambda}_{\testidx}^{\phantom{m}}) &{}- \be\bigl(\mathbf{E}^\mathcal{I}, \boldsymbol{\Lambda}_{\testidx}\bigr)=0,
\label{eq:res_efie}
\\
r_{\mathcal{M}_{\testidx}}(\mathbf{J},\mathbf{I}_m) &{}={}& \ame(\mathbf{J},\boldsymbol{\Lambda}_{\testidx}^m) &{}+{}& \amm(\mathbf{I}_m,\boldsymbol{\Lambda}_{\testidx}^m)&{}=0.
\label{eq:res_slot}
\end{alignat}
Similarly, in terms of~\eqref{eq:res_func_efie} and~\eqref{eq:res_func_slot}, the discretized problems in~\eqref{eq:proj_disc_efie} and~\eqref{eq:proj_disc_slot} are
\begin{alignat}{7}
r_{\mathcal{E}_{\testidx}}(\mathbf{J}_h,\mathbf{I}_h) &{}={}& \aee(\mathbf{J}_h,\boldsymbol{\Lambda}_{\testidx}^{\phantom{m}}) &{}+{}& \aem(\mathbf{I}_h,\boldsymbol{\Lambda}_{\testidx}^{\phantom{m}}) &{}- \be\bigl(\mathbf{E}^\mathcal{I}, \boldsymbol{\Lambda}_{\testidx}\bigr) = 0,
\label{eq:res_disc_efie}
\\
r_{\mathcal{M}_{\testidx}}(\mathbf{J}_h,\mathbf{I}_h) &{}={}& \ame(\mathbf{J}_h,\boldsymbol{\Lambda}_{\testidx}^m) &{}+{}& \amm(\mathbf{I}_h,\boldsymbol{\Lambda}_{\testidx}^m)&{}=0.
\label{eq:res_disc_slot}
\end{alignat}

Through the method of manufactured solutions, \eqref{eq:res_disc_efie} and~\eqref{eq:res_disc_slot} become
\begin{alignat}{7}
r_{\mathcal{E}_{\testidx}}(\mathbf{J}_h,\mathbf{I}_h) &{}={}& r_{\mathcal{E}_{\testidx}}(\mathbf{J}_\text{MS},\mathbf{I}_\text{MS}),
\label{eq:mms_efie}
\\
r_{\mathcal{M}_{\testidx}}(\mathbf{J}_h,\mathbf{I}_h) &{}={}& r_{\mathcal{M}_{\testidx}}(\mathbf{J}_\text{MS},\mathbf{I}_\text{MS}),
\label{eq:mms_slot}
\end{alignat}
where the manufactured solutions are denoted by $\mathbf{J}_\text{MS}$ and $\mathbf{I}_\text{MS}$, and we evaluate $\mathbf{r}_\mathcal{E}(\mathbf{J}_\text{MS},\mathbf{I}_\text{MS})$ and $\mathbf{r}_\mathcal{M}(\mathbf{J}_\text{MS},\mathbf{I}_\text{MS})$ exactly.

As described for the EFIE in~\cite{freno_efie_slot_2024}, rather than solving~\eqref{eq:mms_efie}, we solve~\eqref{eq:proj_disc_efie} by manufacturing the incident electric field $\mathbf{E}^\mathcal{I}$, precluding the need for a dedicated manufactured source term.  Furthermore, we employ the manufactured Green's function~\cite{freno_em_mms_2020,freno_mfie_2022,freno_efie_slot_2024}
\begin{align}
G_\text{MS}(\mathbf{x},\mathbf{x}') = G_q(\mathbf{x},\mathbf{x}') = G_0\biggl(1 - \frac{R^2}{R_m^2}\biggr)^q
\label{eq:G_mms}
\end{align}
to exactly evaluate integrals and prevent contamination from inexact integration in convergence studies.

For the slot equation, inserting~\eqref{eq:res_slot} and~\eqref{eq:res_disc_slot} into~\eqref{eq:mms_slot} yields
\begin{align}
\ame(\mathbf{J}_h,\boldsymbol{\Lambda}_{\testidx}^m) + \amm(\mathbf{I}_h,\boldsymbol{\Lambda}_{\testidx}^m)= \ame(\mathbf{J}_\text{MS},\boldsymbol{\Lambda}_{\testidx}^m) + \amm(\mathbf{I}_\text{MS},\boldsymbol{\Lambda}_{\testidx}^m).
\label{eq:proj_disc_mms_slot}
\end{align}
Rather than solving~\eqref{eq:proj_disc_mms_slot}, we solve~\eqref{eq:proj_disc_slot} for $\mathbf{I}_\text{MS}$ that satisfies
\begin{align}
\ame(\mathbf{J}_\text{MS},\boldsymbol{\Lambda}_{\testidx}^m) + \amm(\mathbf{I}_\text{MS},\boldsymbol{\Lambda}_{\testidx}^m)=0
\label{eq:no_mms_source}
\end{align}
for a given $\mathbf{J}_\text{MS}$.
Consequently, the need for a manufactured source term is precluded for the slot equation as well.
For a known $\mathbf{J}$, as is the case with $\mathbf{J}_\text{MS}$, $\mathbf{I}_m(s)=I_m(s)\mathbf{s}$ can be computing by solving~\eqref{eq:H_slot}.
Projecting~\eqref{eq:H_slot} at the inlet and outlet of the slot onto $\mathbf{s}$ yields
\begin{align}
r^-(s) ={}& J_s^-(s) + \frac{j\omega\epsilon}{2 w L \bigl(k^2 -\beta_x^2\bigr)} \sum_{\modeidx=1}^\infty
\beta_{y_\modeidx}\int_{0}^{\lslot}\sin \biggl(\frac{\modeidx\pi s}{\lslot}\biggr)\sin \biggl(\frac{\modeidx\pi s'}{\lslot}\biggr)\times \nonumber
\\
&
\bigl(
\bigl[I_m^+(s') - I_m^-(s')\bigr] \tan(\beta_{y_\modeidx} d/2) + 
\bigl[I_m^+(s') + I_m^-(s')\bigr] \cot(\beta_{y_\modeidx} d/2)
\bigr)ds' = 0,
\label{eq:slot_ext}
\\
r^+(s) ={}& J_s^+(s) + \frac{j\omega\epsilon}{2 w L \bigl(k^2 -\beta_x^2\bigr)} \sum_{\modeidx=1}^\infty
\beta_{y_\modeidx}\int_{0}^{\lslot}\sin \biggl(\frac{\modeidx\pi s}{\lslot}\biggr)\sin \biggl(\frac{\modeidx\pi s'}{\lslot}\biggr)\times \nonumber
\\
&
\bigl(
\bigl[I_m^-(s') - I_m^+(s')\bigr] \tan(\beta_{y_\modeidx} d/2) + 
\bigl[I_m^+(s') + I_m^-(s')\bigr] \cot(\beta_{y_\modeidx} d/2)
\bigl)ds' = 0,
\label{eq:slot_int}
\end{align}
where $J_s = (\mathbf{J}\times\mathbf{n})\cdot\mathbf{s}$.
Adding~\eqref{eq:slot_ext} and~\eqref{eq:slot_int} yields
\begin{align}
J_s^+(s) + J_s^-(s) + \frac{j\omega\epsilon}{w L \bigl(k^2 -\beta_x^2\bigr)} \sum_{\modeidx=1}^\infty
\beta_{y_\modeidx}\int_{0}^{\lslot}\sin \biggl(\frac{\modeidx\pi s}{\lslot}\biggr)\sin \biggl(\frac{\modeidx\pi s'}{\lslot}\biggr)
\bigl[I_m^+(s') + I_m^-(s')\bigr]   \cot(\beta_{y_\modeidx} d/2)
ds' = 0.
\label{eq:add}
\end{align}
Subtracting~\eqref{eq:slot_ext} from~\eqref{eq:slot_int} yields
\begin{align}
J_s^+(s) - J_s^-(s) + \frac{j\omega\epsilon}{ w L \bigl(k^2 -\beta_x^2\bigr)} \sum_{\modeidx=1}^\infty
\beta_{y_\modeidx}\int_{0}^{\lslot}\sin \biggl(\frac{\modeidx\pi s}{\lslot}\biggr)\sin \biggl(\frac{\modeidx\pi s'}{\lslot}\biggr)
 \bigl[I_m^-(s') - I_m^+(s')\bigr]    \tan(\beta_{y_\modeidx} d/2) 
ds' = 0.
\label{eq:subtract}
\end{align}
We express $J_s(s)$ and $I_m(s)$ as Fourier sine series
\begin{align}
J_s(s) = \sum_{\modeidxp=1}^\infty J_{s_\modeidxp}\sin \biggl(\frac{\modeidxp\pi s}{\lslot}\biggr), \qquad I_m(s) = \sum_{\modeidxp=1}^\infty I_{m_\modeidxp}\sin \biggl(\frac{\modeidxp\pi s}{\lslot}\biggr),
\label{eq:sine_series}
\end{align}
where
\begin{align}
J_{s_\modeidxp} = \frac{2}{\lslot}\int_0^\lslot J_s(s)\sin \biggl(\frac{\modeidxp\pi s}{\lslot}\biggr) ds.
\label{eq:jsq}
\end{align}
To obtain the coefficients $I_{m_\modeidxp}$~\eqref{eq:sine_series}, we insert $J_s(s)$ and $I_m(s)$~\eqref{eq:sine_series} into~\eqref{eq:add} and~\eqref{eq:subtract} and account for orthogonality:
\begin{align}
I_{m_\modeidxp}^- &{}= \frac{j w \bigl(k^2 -\beta_x^2\bigr)}{\beta_{y_\modeidxp}\omega\epsilon} \bigl(\bigl[J_{s_\modeidxp}^+ + J_{s_\modeidxp}^-\bigr]\tan(\beta_{y_\modeidxp} d/2) + \bigl[J_{s_\modeidxp}^+ - J_{s_\modeidxp}^-\bigr]\cot(\beta_{y_\modeidxp} d/2)\bigr), \label{eq:imqm} \\
I_{m_\modeidxp}^+ &{}= \frac{j w \bigl(k^2 -\beta_x^2\bigr)}{\beta_{y_\modeidxp}\omega\epsilon} \bigl(\bigl[J_{s_\modeidxp}^+ + J_{s_\modeidxp}^-\bigr]\tan(\beta_{y_\modeidxp} d/2) - \bigl[J_{s_\modeidxp}^+ - J_{s_\modeidxp}^-\bigr]\cot(\beta_{y_\modeidxp} d/2)\bigr).
\label{eq:imqp}
\end{align}
With~\eqref{eq:imqm} and~\eqref{eq:imqp}, $I_m(s)$~\eqref{eq:sine_series} is known.

\subsection{Solution-Discretization Error} 
\label{sec:sde}

The solution-discretization error is the result of the basis-function approximations to the solutions~\eqref{eq:sol_h}.  We measure this error from the discretization errors 
\begin{alignat}{7}
&\mathbf{e}_\mathbf{J} &&{}={}& \mathbf{J}^h &{}-{}& \mathbf{J}&_n&,
\label{eq:solution_error_J}
\\
&\mathbf{e}_\mathbf{I} &&{}={}& \mathbf{I}^h &{}-{}& \mathbf{I}&_s&,
\label{eq:solution_error_I}
\end{alignat}
where $J_{n_\srcidx}$ is the component of $\mathbf{J}_\text{MS}$ that flows across triangle pair $\srcidx$, and $I_{s_\srcidx}$ is the component of $\mathbf{I}_\text{MS}$ that flows across one-dimensional element pair $\srcidx$.  Letting $h$ indicate mesh size, we expect the norms of~\eqref{eq:solution_error_J} and~\eqref{eq:solution_error_I} to be $\mathcal{O}(h^2)$.

However, as described in~\cite{freno_efie_slot_2024}, the presence of a wire on a surface yields a discontinuity, which appears in the first term in $\aem(\mathbf{u},\mathbf{v})$~\eqref{eq:aem}.  For the surface, such a discontinuity reduces the convergence rate to $\mathcal{O}(h)$~\cite{dangelo_2012,li_2021}.  To prevent the discontinuity from contaminating convergence studies, we can remove the discontinuity from the submatrix $\mathbf{B}$ in $\mathbf{Z}$~\eqref{eq:Z} using the $\mathbf{C}$ submatrix, with the corresponding contribution to the manufactured incident electric field $\mathbf{E}^\mathcal{I}$ being removed as well~\cite{freno_efie_slot_2024}.

\subsection{Numerical-Integration Error} 
\label{sec:nie}

The numerical-integration error is the result of the generally approximate quadrature evaluations of the integrals in~\eqref{eq:proj_disc_efie} and~\eqref{eq:proj_disc_slot}.  We measure this error using~\cite{freno_mfie_2022}
\begin{alignat}{7}
&e_a&{}={}&
\boldsymbol{\mathcal{J}}^H(&\mathbf{Z}^q{}-{}&\mathbf{Z}&&)\boldsymbol{\mathcal{J}},  \label{eq:a_error_cancel}
\\
&e_b&{}={}&
\boldsymbol{\mathcal{J}}^H(&\mathbf{V}^q{}-{}&\mathbf{V}&&), \label{eq:b_error_cancel}
\end{alignat}
where $\mathbf{Z}^q$ and $\mathbf{V}^q$ denote the quadrature evaluations of $\mathbf{Z}$ and $\mathbf{V}$, and $\mathbf{Z}$ and $\mathbf{V}$ are evaluated exactly in~\eqref{eq:a_error_cancel} and~\eqref{eq:b_error_cancel}.  Additionally,
\begin{align*}
\boldsymbol{\mathcal{J}} = \left\{\begin{array}{@{} r @{} l @{}}
\mathbf{J}&_n \\ 
\mathbf{I}&_s\end{array}\right\}.
\end{align*}
Equations~\eqref{eq:a_error_cancel} and~\eqref{eq:b_error_cancel} avoid contamination from the solution-discretization error.
The absolute values of~\eqref{eq:a_error_cancel} and~\eqref{eq:b_error_cancel} are expected to be $\mathcal{O}(h^p)$, where $p$ depends on the quadrature accuracy.

\subsection{Series Truncation for $I_m(s)$}
\label{sec:trunc}

To tractably accommodate $I_m(s)$, it is necessary to truncate the infinite series representation~\eqref{eq:sine_series}.  In this subsection, we show the convergence implications of the truncation.

\subsubsection{$J_s(s)$}
\label{sec:js}

We begin by considering the convergence of the truncation error for $J_s(s)$. Let 
\begin{align*}
J_{s_{\modemaxp}}(s) = \sum_{\modeidxp=1}^\modemaxp J_{s_{\modeidxp}}\sin \biggl(\frac{\modeidxp\pi s}{\lslot}\biggr)
\end{align*}
denote the truncation of the infinite series representation for $J_s(s)$~\eqref{eq:sine_series} and 
\begin{align*}
e_{J_{\modemaxp}}(s) =  J_{s_{\modemaxp}}(s) -J_s(s) = - \!\!\!\sum_{\modeidxp=\modemaxp+1}^\infty\!\!\! J_{s_{\modeidxp}}\sin \biggl(\frac{\modeidxp\pi s}{\lslot}\biggr)
\end{align*}
denote the difference between the truncated and infinite series representations.  Additionally, let
\begin{align}
\bigl\|e_{J_{\modemaxp}}(s)\bigr\|_\infty = \max_{s\in[0,\,\lslot]} \bigl|e_{J_{\modemaxp}}(s)\bigr|.
\label{eq:eJQ_norm}
\end{align}
We note that $|\sin (\modeidxp\pi s/\lslot)|\le 1$,
such that, in~\eqref{eq:eJQ_norm},
\begin{align}
\bigl|e_{J_{\modemaxp}}(s)\bigr| \le \!\!\sum_{\modeidxp=\modemaxp+1}^\infty \!\!\bigl|J_{s_{\modeidxp}}\bigr|.
\label{eq:jineq1}
\end{align}
For a sufficiently large $\modeidxp$,
\begin{align}
\bigl|J_{s_{\modeidxp}}\bigr|\le C_{J_{\modeidxp}}{\modeidxp}^{-p},
\label{eq:jsq_limit}
\end{align}
where $p$ is determined from the manufactured $J_s(s)$.
Assuming $p>1$, in~\eqref{eq:jineq1}, 
\begin{align*}
\sum_{\modeidxp=\modemaxp+1}^\infty\!\! \bigl|J_{s_{\modeidxp}}\bigr| \le C_{J_{\infty}}\!\!\!\sum_{\modeidxp=\modemaxp+1}^\infty\!\!\! {\modeidxp}^{-p} \approx C_{J_{\infty}}\int_{\modemaxp+1}^\infty {\modeidxp}^{-p} d{\modeidxp} = \frac{C_{J_{\infty}}}{p-1} (\modemaxp+1)^{1-p},
\end{align*}
where $C_{J_{\infty}}$ is an upper bound for $C_{J_{\modeidxp}}$.
Consequently, $\|e_{J_{\modemaxp}}(s)\|_\infty$ is $\mathcal{O}({\modemaxp}^{1-p})$.

\subsubsection{$I_m(s)$}
\label{sec:im}

Next, we consider the convergence of the truncation error for $I_m(s)$.
As shown in~\eqref{eq:imqm} and~\eqref{eq:imqp}, $I_{m_{\modeidxp}}$ is related to ${\modeidxp}$ through a linear combination of $[J_{s_{\modeidxp}}^+ + J_{s_{\modeidxp}}^-]\tan(\beta_{y_{\modeidxp}} d/2)/\beta_{y_{\modeidxp}}$ and $[J_{s_{\modeidxp}}^+ - J_{s_{\modeidxp}}^-]\cot(\beta_{y_{\modeidxp}} d/2)/\beta_{y_{\modeidxp}}$.   Noting that, from~\eqref{eq:prop_const}, for large values of $\modeidxp$,
\begin{align*}
\beta_{y_{\modeidxp}} \approx j\frac{\modeidxp\pi}{\lslot},
\end{align*}
and 
\begin{alignat*}{7}
\frac{\tan(\beta_{y_{\modeidxp}} d/2)}{\beta_{y_{\modeidxp}}} &{}\approx \phantom{-}\frac{\lslot\tanh(\modeidxp\pi d/(2\lslot))}{\modeidxp\pi} &&{}\approx\phantom{-}\frac{\lslot}{\modeidxp\pi},
\\
\frac{\cot(\beta_{y_{\modeidxp}} d/2)}{\beta_{y_{\modeidxp}}}&{}\approx -\frac{\lslot\coth(\modeidxp\pi d/(2\lslot))}{\modeidxp\pi} &&{}\approx-\frac{\lslot}{\modeidxp\pi},
\end{alignat*}
such that both are $\mathcal{O}({\modeidxp}^{-1})$. From~\eqref{eq:jsq_limit}, $[J_{s_{\modeidxp}}^+ + J_{s_{\modeidxp}}^-]\tan(\beta_{y_{\modeidxp}} d/2)/\beta_{y_{\modeidxp}}$ and $[J_{s_{\modeidxp}}^+ - J_{s_{\modeidxp}}^-]\cot(\beta_{y_{\modeidxp}} d/2)/\beta_{y_{\modeidxp}}$  are both $\mathcal{O}({\modeidxp}^{-p-1})$. 
Consequently, $I_{m_{\modeidxp}}$ is $\mathcal{O}({\modeidxp}^{-p-1})$:
\begin{align*}
\bigl|I_{m_{\modeidxp}}\bigr|\le C_{I_{\modeidxp}}{\modeidxp}^{-p-1}.
\end{align*}
Let 
\begin{align}
I_{m_{\modemaxp}}(s) = \sum_{\modeidxp=1}^\modemaxp I_{m_{\modeidxp}}\sin \biggl(\frac{\modeidxp\pi s}{\lslot}\biggr)
\label{eq:ImQ}
\end{align}
denote the truncation of the infinite series representation for $I_m(s)$~\eqref{eq:sine_series} and
\begin{align}
e_{I_{\modemaxp}}(s) =  I_{m_{\modemaxp}}(s) -I_m(s) = - \!\!\!\sum_{\modeidxp=\modemaxp+1}^\infty\!\!\! I_{m_{\modeidxp}}\sin \biggl(\frac{\modeidxp\pi s}{\lslot}\biggr)
\label{eq:eIq}
\end{align}
denote the difference between the truncated and infinite series representations.  Additionally, let
\begin{align}
\bigl\|e_{I_{\modemaxp}}(s)\bigr\|_\infty = \max_{s\in[0,\,\lslot]} \bigl|e_{I_{\modemaxp}}(s)\bigr|.
\label{eq:eIQ_norm}
\end{align}
In~\eqref{eq:eIQ_norm},
\begin{align*}
\bigl|e_{I_{\modemaxp}}(s)\bigr| \le \!\!\sum_{\modeidxp=\modemaxp+1}^\infty\!\! \bigl|I_{m_{\modeidxp}}\bigr| \le C_{I_{\infty}}\!\!\!\sum_{\modeidxp=\modemaxp+1}^\infty\!\!\! {\modeidxp}^{-p-1} \approx C_{I_{\infty}}\int_{\modemaxp+1}^\infty {\modeidxp}^{-p-1} d{\modeidxp} = \frac{C_{I_{\infty}}}{p} (\modemaxp+1)^{-p},
\end{align*}
where $C_{I_{\infty}}$ is an upper bound for $C_{I_{\modeidxp}}$.
Therefore, $\|e_{I_{\modemaxp}}(s)\|_\infty$ is $\mathcal{O}({\modemaxp}^{-p})$.

\subsubsection{Integration of $I_m(s)$}
\label{sec:im_int}

When integrating $I_{m_{\modemaxp}}$~\eqref{eq:ImQ} over different meshes and increasing $\modemaxp$ with the number of triangles $n_t$, $e_{I_{\modemaxp}}(s)$~\eqref{eq:eIq} introduces an error.  To derive the convergence rate of this error, we begin by considering the integral of the error:
\begin{align*}
\int_0^{\lslot} e_{I_{\modemaxp}}(s) ds 
=  
-\!\!\!\sum_{\modeidxp=\modemaxp+1}^\infty\!\! I_{m_{\modeidxp}}\int_0^{\lslot}\sin \biggl(\frac{\modeidxp\pi s}{\lslot}\biggr) ds 
= 
\frac{\lslot}{\pi}\!\!\sum_{\modeidxp=\modemaxp+1}^\infty\!\! \frac{I_{m_{\modeidxp}}}{\modeidxp}\bigl(-1+(-1)^\modeidxp\bigr),
\end{align*}
which can be bounded by
\begin{align*}
\biggl|\int_0^{\lslot} e_{I_{\modemaxp}}(s) ds \biggr| 
\le 
\frac{2\lslot}{\pi}\!\!\sum_{\modeidxp=\modemaxp+1}^\infty\!\! \frac{|I_{m_{\modeidxp}}|}{\modeidxp} 
\le 
\frac{2\lslot}{\pi}\!\!\sum_{\modeidxp=\modemaxp+1}^\infty\!\! {C_{I_{\modeidxp}}{\modeidxp}^{-p-2}}
\approx 
\frac{2\lslot C_{I_{\infty}}}{\pi}\int_{\modemaxp+1}^\infty {\modeidxp}^{-p-2} d{\modeidxp} 
= 
\frac{2\lslot C_{I_{\infty}}}{(p+1)\pi} (\modemaxp+1)^{-p-1}
.
\end{align*}
Therefore, $|\int_0^{\lslot} e_{I_{\modemaxp}}(s) ds |$ is $\mathcal{O}({\modemaxp}^{-p-1})$.  When assessing the convergence rate of the numerical integration, if $\modemaxp\sim 1/h$, where $1/h\sim\sqrt{n_t}$, and the integration error convergence is faster than $\mathcal{O}(h^{p+1})$, the convergence rate will be limited to $\mathcal{O}(h^{p+1})$. While $\modemaxp$ can be increased faster than $h$, $\modemaxp$ can instead be held constant to avoid this issue.

\section{Numerical Examples} 
\label{sec:results}

In this section, we illustrate the methods outlined in Section~\ref{sec:mms} by separately measuring the solution-discretization error (Section~\ref{sec:sde}) and numerical-integration error (Section~\ref{sec:nie}).

\subsection{Domain and Coordinate Systems} 
\label{sec:domain}

\begin{figure}
\centering
\begin{subfigure}[b]{.99\textwidth}
\centering
\includegraphics[scale=.28,clip=true,trim=0in 0in 0in 0in]{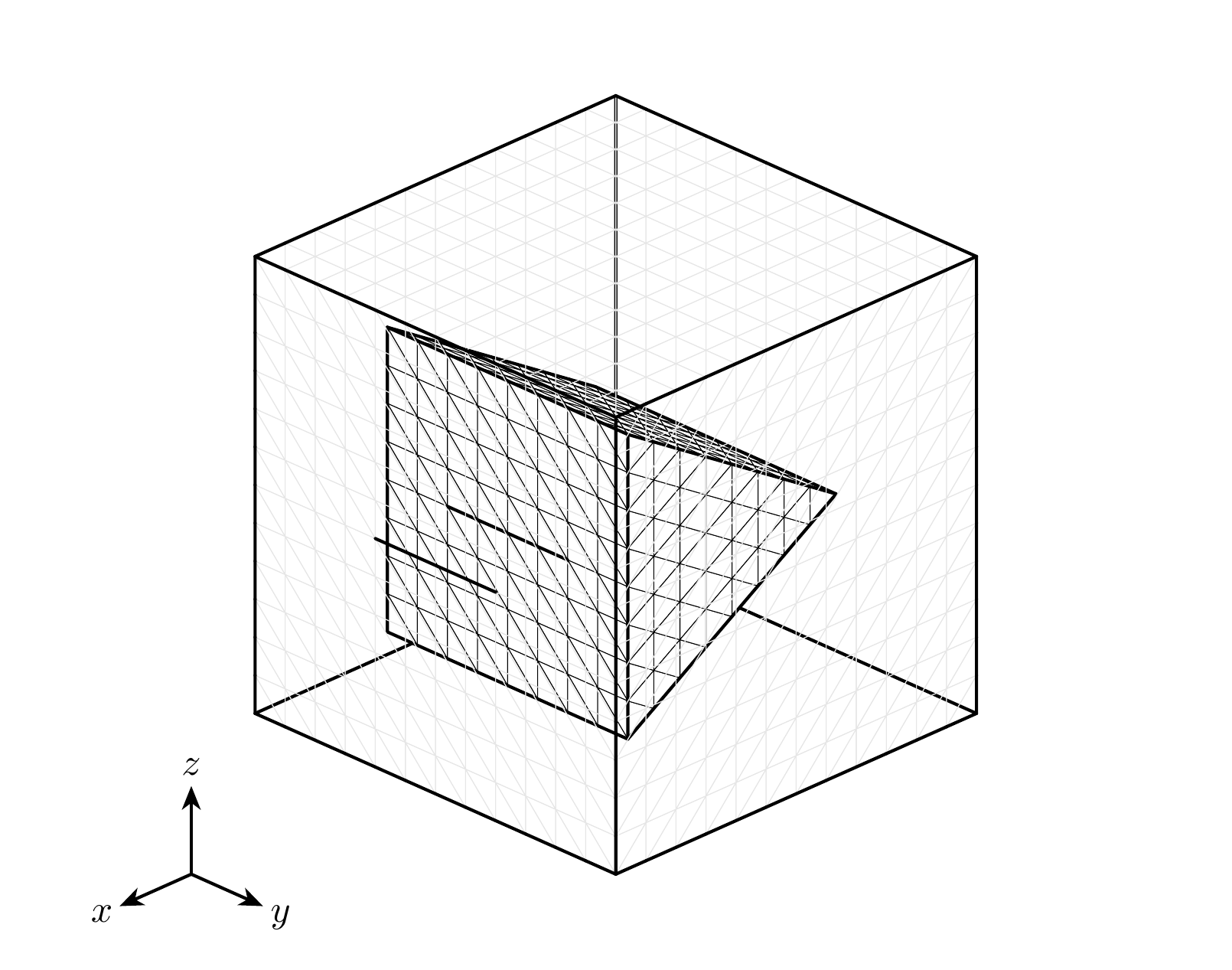}
\includegraphics[scale=.28,clip=true,trim=0in 0in 0in 0in]{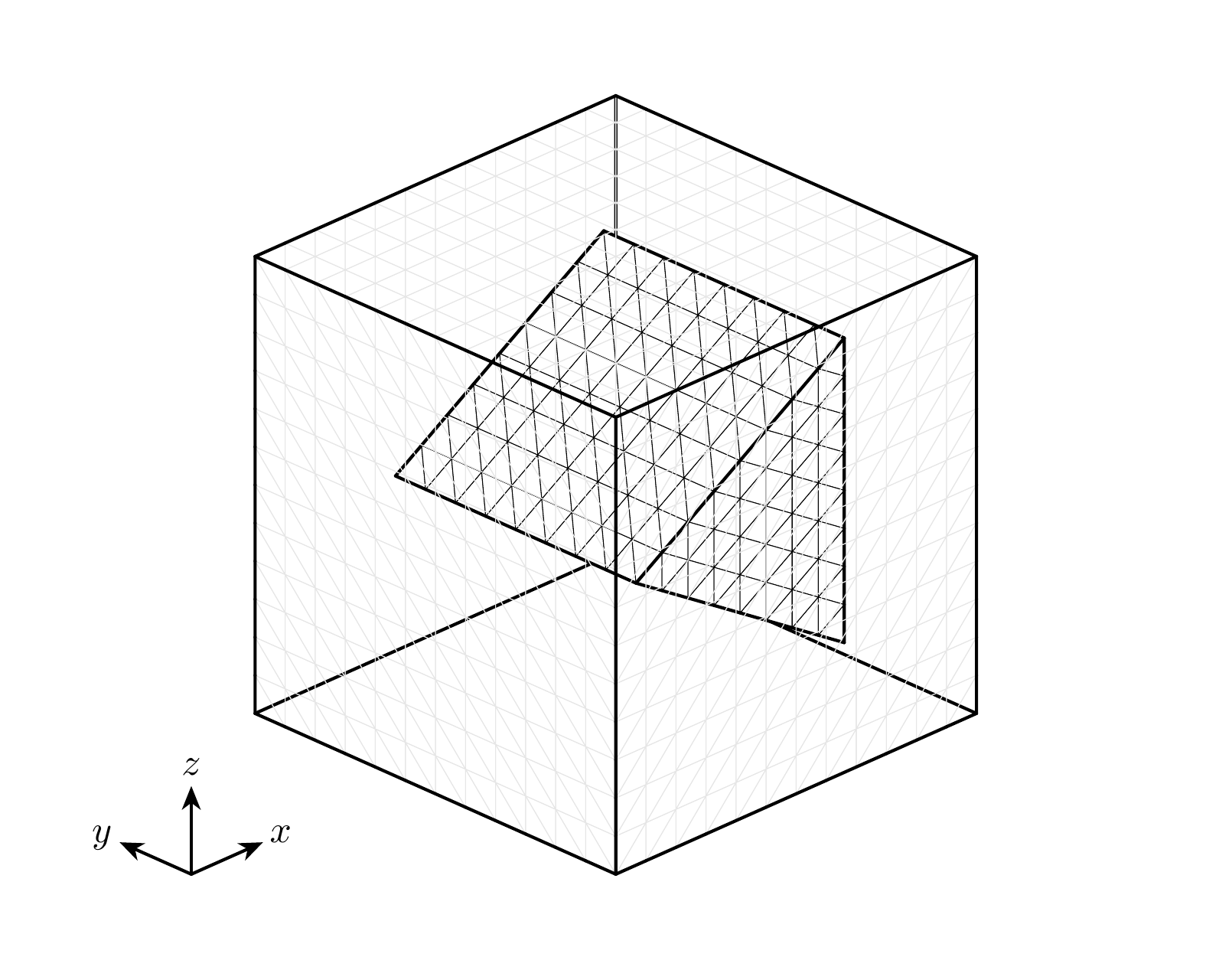}
\caption{$d_1=\lext/5$}
\end{subfigure}
\\
\begin{subfigure}[b]{.99\textwidth}
\centering
\includegraphics[scale=.28,clip=true,trim=0in 0in 0in 0in]{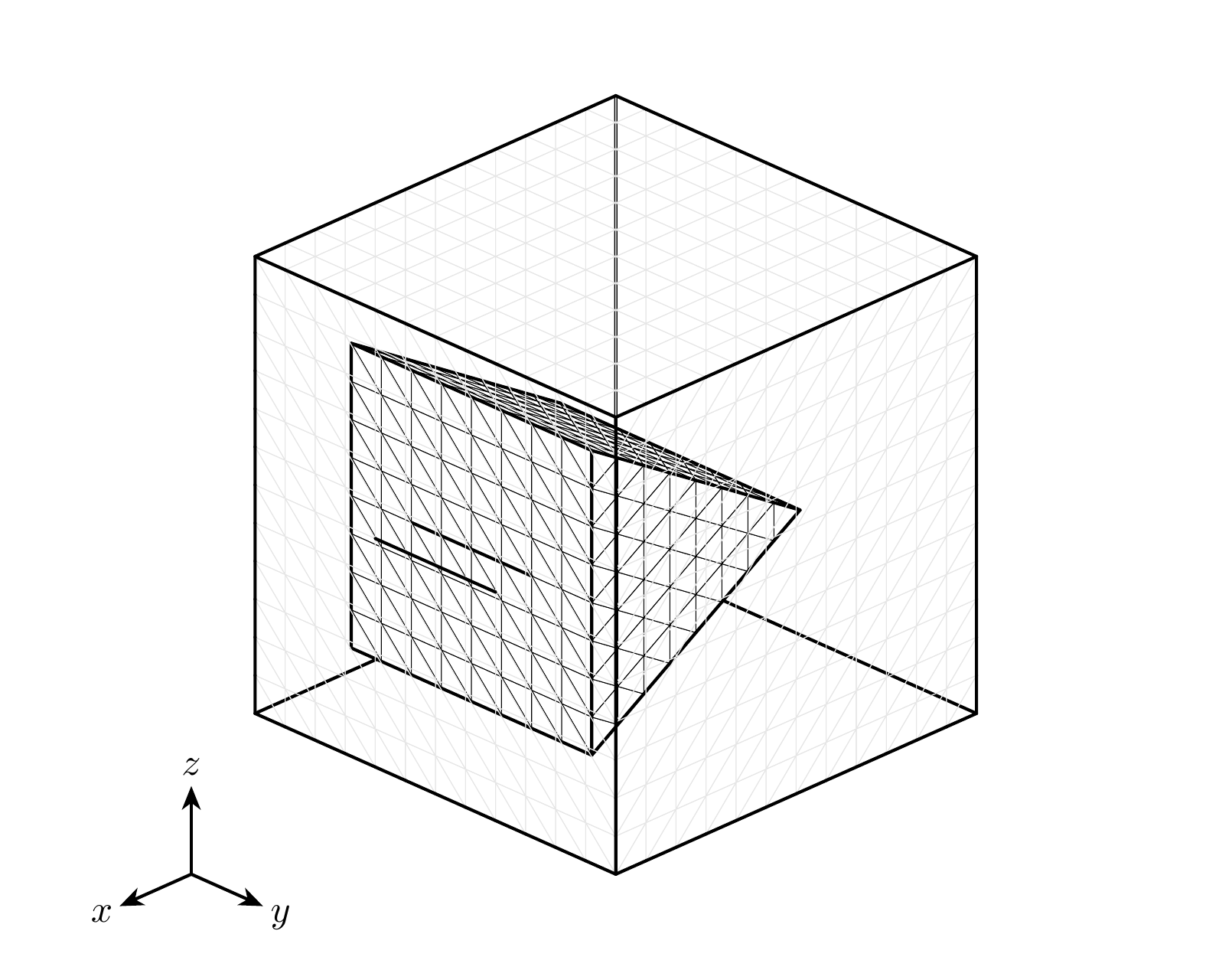}
\includegraphics[scale=.28,clip=true,trim=0in 0in 0in 0in]{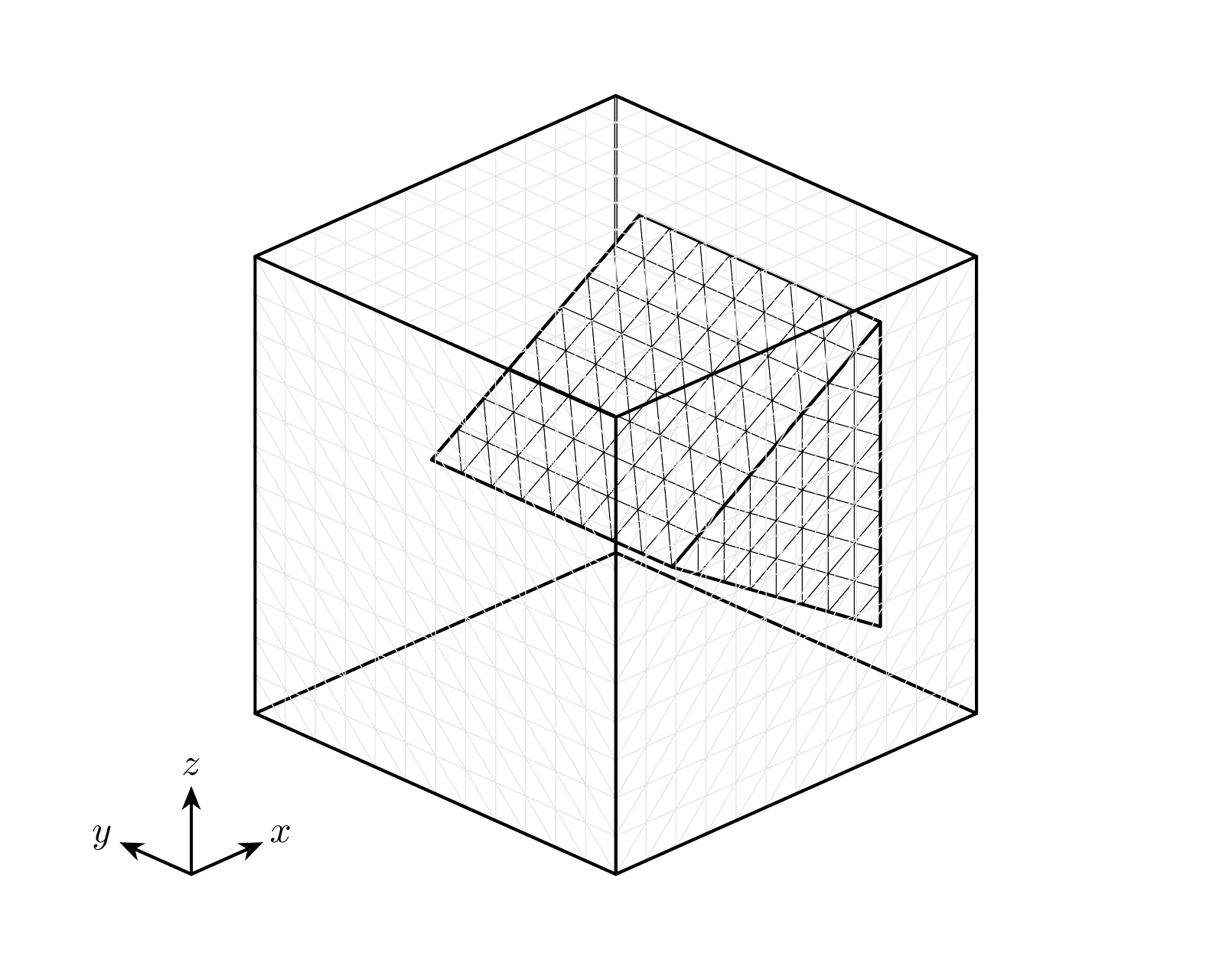}
\caption{$d_2=\lext/10$}
\end{subfigure}
\\
\begin{subfigure}[b]{.99\textwidth}
\centering
\includegraphics[scale=.28,clip=true,trim=0in 0in 0in 0in]{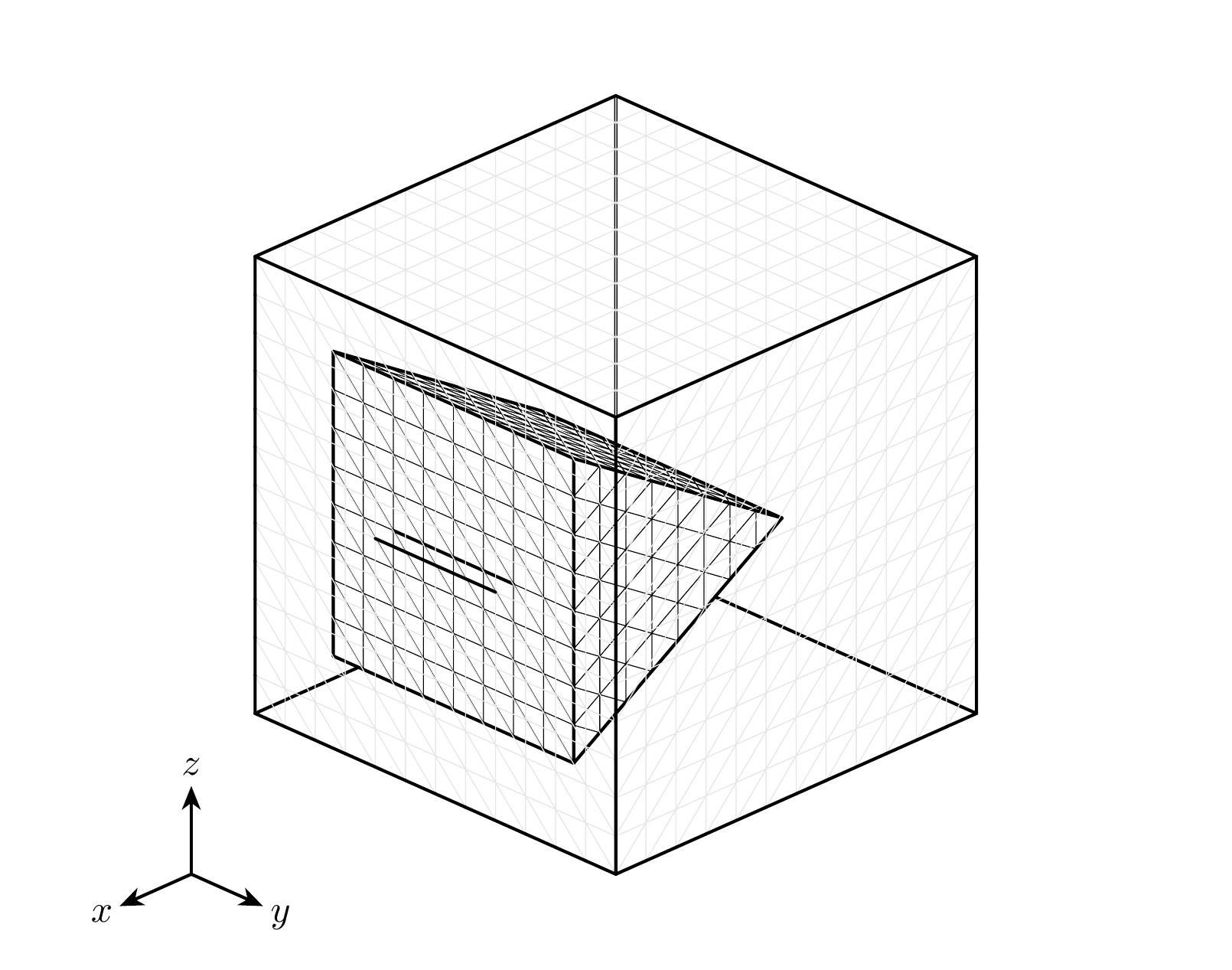}
\includegraphics[scale=.28,clip=true,trim=0in 0in 0in 0in]{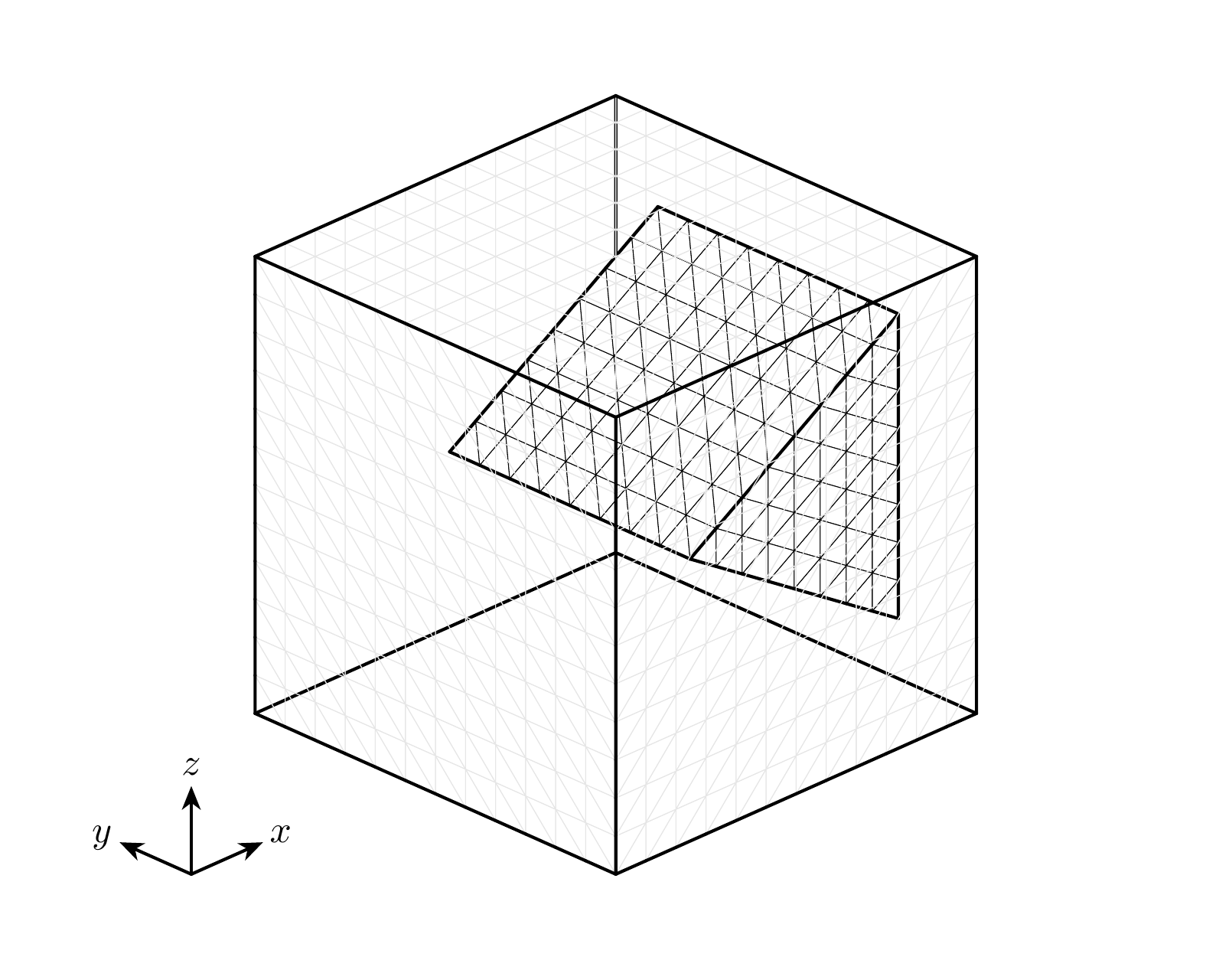}
\caption{$d_3=\lext/20$}
\end{subfigure}
\caption{Discretized domain using $\ntriangles=2240$ triangles for 3 depths.}
\vskip-\dp\strutbox
\label{fig:whole_mesh_12}
\end{figure}

\begin{figure}
\centering
\input{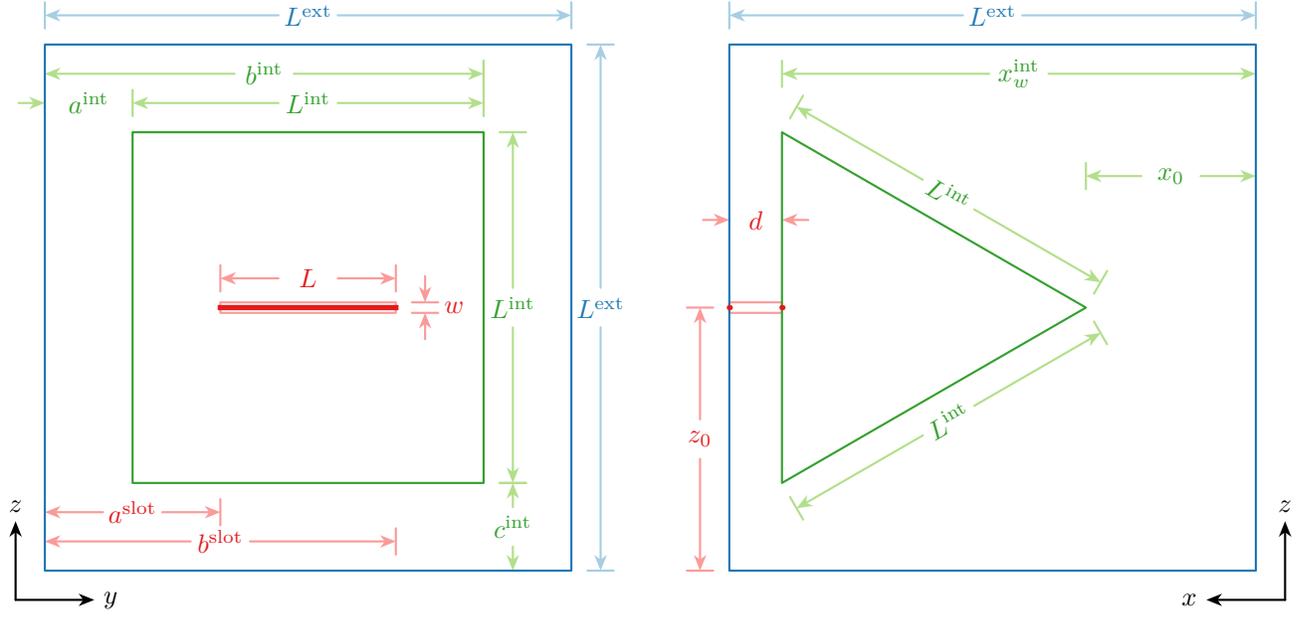}
\caption{Dimensions of the domain, which are specified in~\cite{freno_efie_slot_2024}.}
\vskip-\dp\strutbox
\label{fig:dimensions}
\end{figure}

In general, the method of manufactured solutions allows considerable freedom for selecting the solution, geometry, parameters, and boundary conditions for the manufactured problem.  However, sufficiently smooth solutions are required to measure expected convergence rates, and the interior and exterior surfaces of the scatterer are required to be polyhedra in order to be exactly represented by planar elements.  The presence of the slot places additional constraints on the behavior of the solution in the vicinity of the slot.  Furthermore, while geometries and solutions of arbitrary complexity may be considered, additional complexity 
will generally incur additional computational expense.  \rereading{Other types of testing, such as solution-verification and regression tests, are more appropriate for complex geometries and solutions, and should be used in addition to the code-verification approaches discussed here.  Therefore, in the context of code verification, we seek geometries and solutions that are simple, yet nontrivial.}
For this work, we consider the scatterer geometry shown in Figures~\ref{fig:whole_mesh_12} and~\ref{fig:dimensions} and introduced in~\cite{freno_efie_slot_2024}.  The exterior surface is a cube, and the surface bounding the interior cavity is a triangular prism.  The interior and exterior field domains are connected by a rectangularly prismatic slot.  The slot is modeled by a pair of wires, with one positioned along each opening.  

For this problem, we consider three depths: $d_1=\lext/5$, $d_2=\lext/10$, and $d_3=\lext/20$, as well as two Green's functions~\eqref{eq:G_mms}: $G_1$ and $G_2$. 
For each of the three depths, Figure~\ref{fig:whole_mesh_12} shows the discretized domains using $\ntriangles=2240$ for the surfaces and four one-dimensional bar elements for each wire.
For the medium that surrounds the scatterer exterior and occupies the cavity interior, we set the permeability and permittivity to those of free space: $\mu =\mu_0$ and $\epsilon = \epsilon_0$, assuming zero electrical conductivity ($\sigma=0$), and we choose a wavenumber of $k=2\pi$~m$^{-1}$.  We set the electrical conductivity of the scatterer to that of aluminum.  For the medium that occupies the slot interior, we set $\mu =\mu_0$ and $\sigma=5$~S/m, such that the medium is characterized by a complex permittivity 
\begin{align}
\epsilon = \epsilon' - j\frac{\sigma}{\omega},
\label{eq:comp_perm}
\end{align}
for which we set $\epsilon' = \epsilon_0$. 

\begin{figure}
\centering
\includegraphics[scale=.28,clip=true,trim=0in 0in 0in 0in]{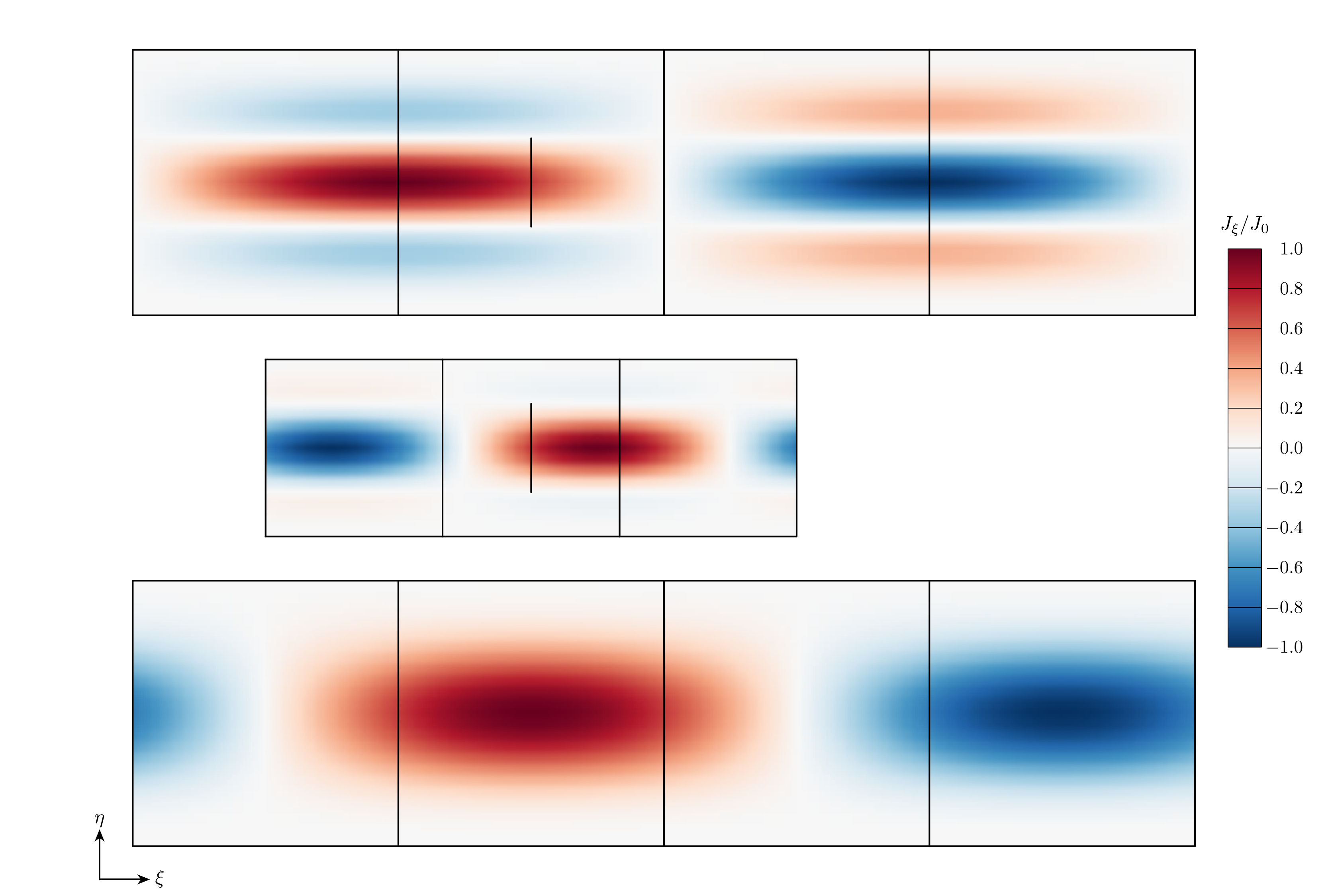}
\caption{\strut Components of $\mathbf{J}_\text{MS}$: $J_{\xi_\theta}$~\eqref{eq:j_xi_theta} for the scatterer (top) and cavity (middle), and $J_{\xi_\phi}$~\eqref{eq:j_xi_phi} for the scatterer (bottom).}
\vskip-\dp\strutbox
\label{fig:2d_solutions}
\end{figure}

\begin{figure}
\centering
\includegraphics[scale=.28,clip=true,trim=0in 0in 0in 0in]{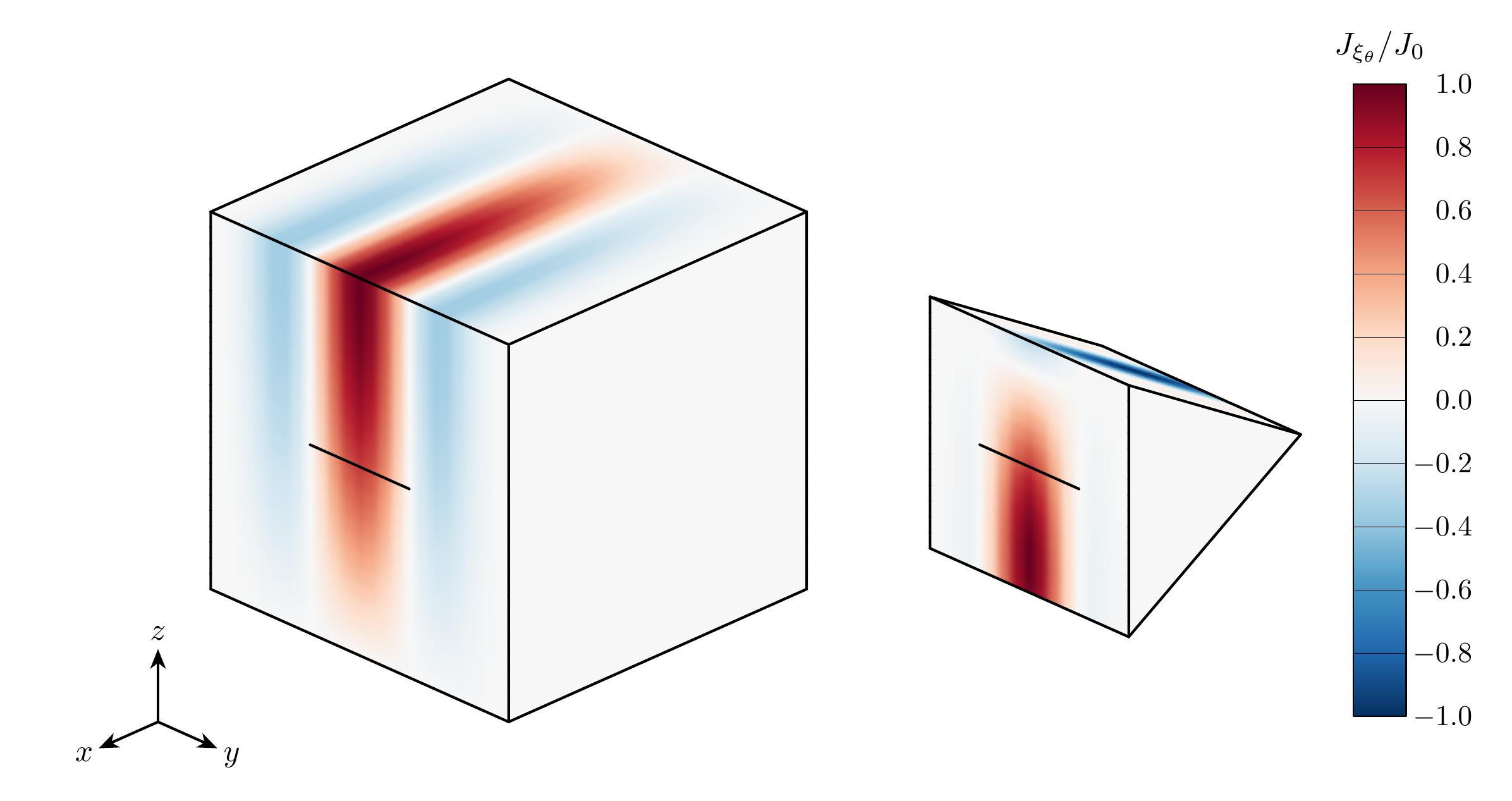}
\includegraphics[scale=.28,clip=true,trim=0in 0in 0in 0in]{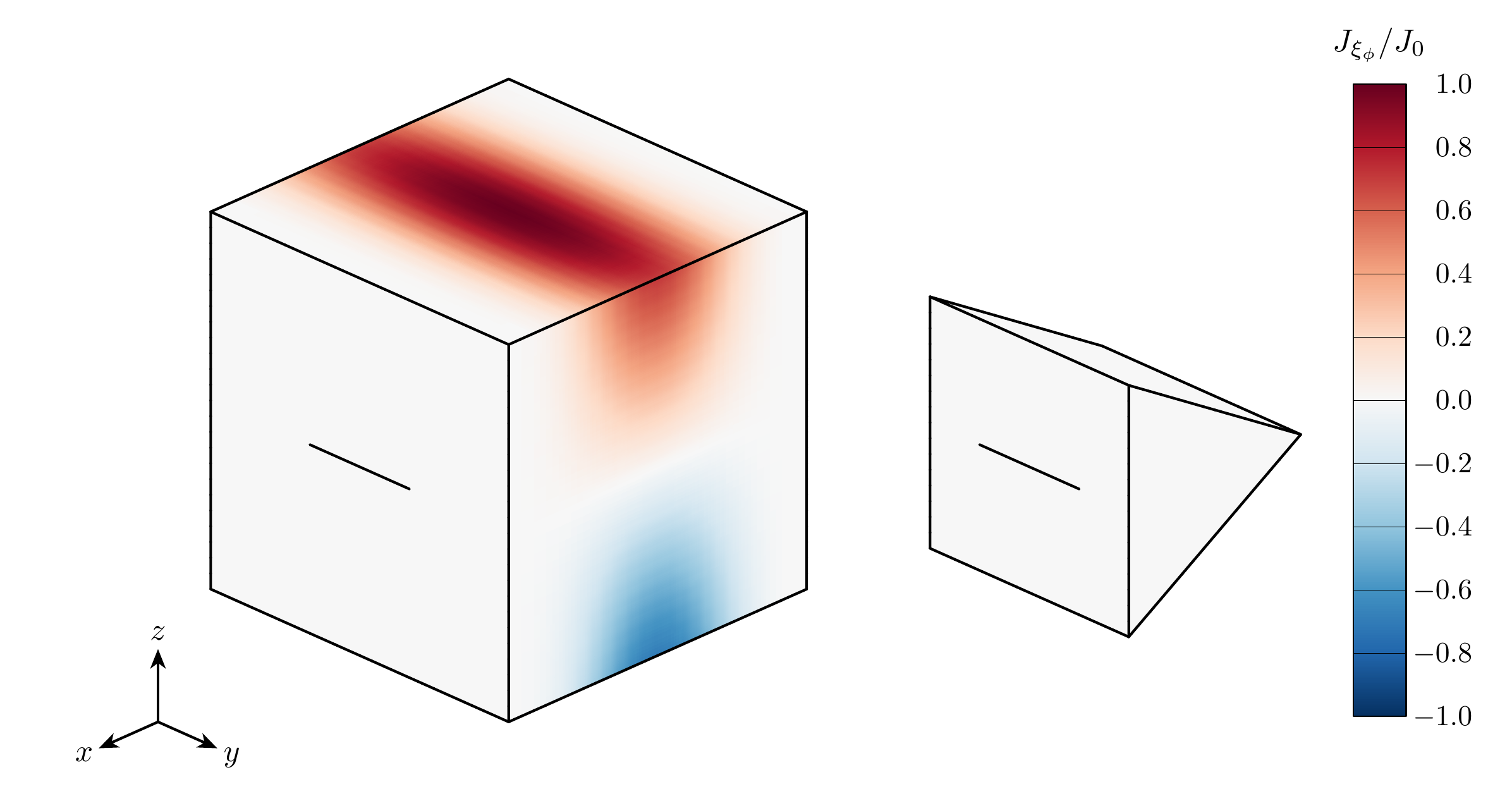}
\includegraphics[scale=.28,clip=true,trim=0in 0in 0in 0in]{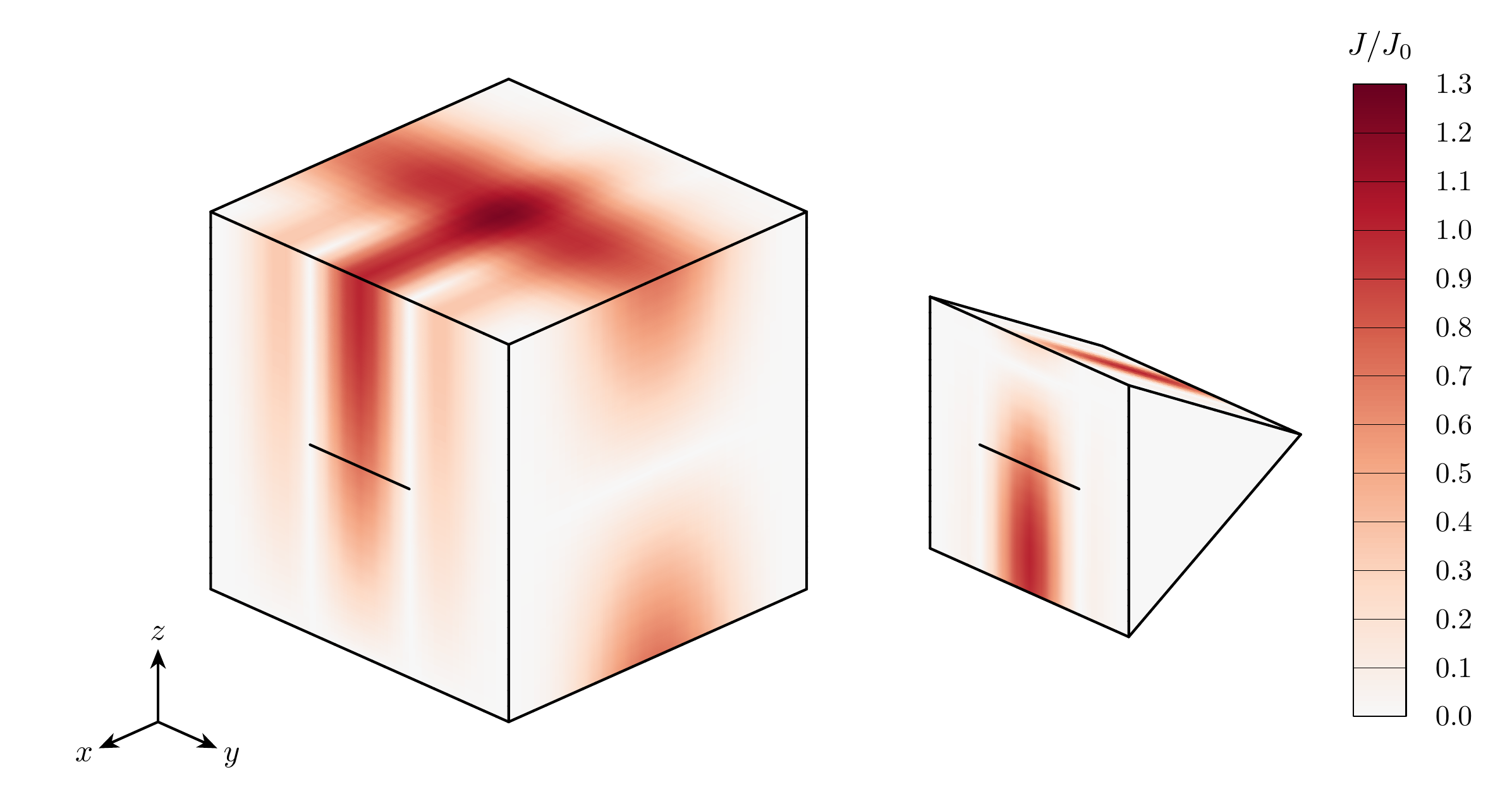}
\caption{\strut Components of $\mathbf{J}_\text{MS}$: $J_{\xi_\theta}$~\eqref{eq:j_xi_theta} (top), $J_{\xi_\phi}$~\eqref{eq:j_xi_phi} (middle), and $J=|\mathbf{J}_\text{MS}|$~\eqref{eq:J_MS_cube} and~\eqref{eq:J_MS_triangular_prism} (bottom).}
\vskip-\dp\strutbox
\label{fig:3d_solutions}
\end{figure}

When manufacturing the surface current, we employ coordinate systems that conform to the lateral surfaces of the cube and triangular prism~\cite{freno_efie_slot_2024}.  
For both geometries, we use $\boldsymbol{\xi}_\theta$, for which $\eta=y$ and $\xi$ is perpendicular to $y$, wrapping counterclockwise (per the right-hand rule) around $y$ along the surfaces where $\mathbf{n}\cdot\mathbf{e}_y=0$.
For the cubic scatterer, $\eta\in[0,\,1]\lext$ and $\xi\in[0  ,\,  4] \lext$.
For the triangularly prismatic cavity, $\eta\in[\aint,\,\bint]$ and $\xi\in\xi_0+[0,\,3]\lint$, where $\xi_0=3(\lext-\lint)/2$.  For both geometries, the wires are positioned at $\xi_w=3\lext/2$ for $\eta\in[\aslot,\bslot]$.
For the cubic scatterer, we additionally use $\boldsymbol{\xi}_\phi$, for which $\eta=x$ and $\xi$ is perpendicular to $x$, wrapping counterclockwise around $x$ along the surfaces where $\mathbf{n}\cdot\mathbf{e}_x=0$.
For $\boldsymbol{\xi}_\phi$, $\eta\in[0,\,1]\lext$ and $\xi\in[0  ,\,  4] \lext$.

\subsection{Manufactured Surface Current} 

For the cube, we prescribe the form of the manufactured surface current density as
\begin{align}
\mathbf{J}_\text{MS}(\mathbf{x}) = J_{\xi_\theta}(\boldsymbol{\xi}_\theta)\mathbf{e}_{\xi_\theta} + J_{\xi_\phi}(\boldsymbol{\xi}_\phi)\mathbf{e}_{\xi_\phi},
\label{eq:J_MS_cube}
\end{align}
and we prescribe the form for the triangular prism as 
\begin{align}
\mathbf{J}_\text{MS}(\mathbf{x}) = J_{\xi_\theta}(\boldsymbol{\xi}_\theta)\mathbf{e}_{\xi_\theta},
\label{eq:J_MS_triangular_prism}
\end{align}
where we express the dependencies of the magnitudes as separable:
\begin{align}
J_{\xi_\theta}(\boldsymbol{\xi}) &{}= J_0 f_{\xi_\theta}(\xi)g_{\eta_\theta}(\eta), \label{eq:j_xi_theta}
\\
J_{\xi_\phi}  (\boldsymbol{\xi}) &{}= J_0 f_{\xi_\phi}  (\xi)g_{\eta_\phi}(\eta), \label{eq:j_xi_phi}
\end{align}
with $J_0=1$~A/m.

For $f_\xi(\xi)$ and $g_\eta(\eta)$, we opt for nontrivial functions that are at least of class $C^2$.  To avoid the need for finer meshes in the mesh-convergence studies, we seek to minimize oscillations.  For $f_\xi(\xi)$, we use periodic functions with a single period over the domain:
\begin{align*}
f_{\xi_\theta}(\xi)&{}=\sin(\gamma(\xi-\bar{\xi}_1)), \\
f_{\xi_\phi}(\xi)&{}=\sin(\gamma(\xi-\bar{\xi}_2)).
\end{align*}
We choose $\gamma=\pi/(2\lext)$, $\bar{\xi}_1=0$, and $\bar{\xi}_2=\lext/2$ for the cube, and we choose $\gamma=2\pi/(3\lint)$ and $\bar{\xi}_1=5\lext/4$ for the triangular prism.
For $g_{\eta_\phi}(\eta)$, 
\begin{align*}
g_{\eta_\phi}(\eta)&{}=\sin^3\biggl(\frac{\pi\eta}{\lext}\biggr)
\end{align*}
results in $g_{\eta_\phi}(\eta)$ and its first and second derivatives being zero at $\eta=\{0,\,\lext\}$, such that it is of class $C^2$ and is therefore suitable.  
Similarly, $g_{\eta_\theta}(\eta)$ is of class $C^2$ if $g_{\eta_\theta}(\eta)$ is $C^\infty$ for $\eta\in(a,\,b)$ and $g_{\eta_\theta}(\eta)$ and its first and second derivatives are zero at $\eta=\{a,\,b\}$.  Additionally, from~\eqref{eq:sine_series}, $g_{\eta_\theta}(\eta)$ must be zero at $s=\{0,\,\lslot\}$ ($\eta=\{\aslot,\,\bslot\}$).  This additional constraint does not exist for the thick slot model~\cite{freno_efie_slot_2024}.
Therefore, we choose
\begin{align}
g_{\eta_\theta}(\eta) = \sum_{\modeidxp=1}^3 C_{\modeidxp}\sin\bigg(\modeidxp'\pi\frac{\eta-a}{b-a}\bigg),
\label{eq:g_eta_theta}
\end{align}
where $\modeidxp'=2\modeidxp-1$, to minimize oscillations.  In~\eqref{eq:g_eta_theta}, for the cube, $a=\aext=0$ and $b=\bext=\lext$; for the triangular prism, $a=\aint$ and $b=\bint$.  We set $C^\text{ext}=\{1/4,\,-1/2,\,1/4\}$ for the cube and $C^\text{int}=\{1/2,\,-3/8,\,1/8\}$ for the triangular prism.
Figures~\ref{fig:2d_solutions} and~\ref{fig:3d_solutions} show $J_{\xi_\theta}(\boldsymbol{\xi})$~\eqref{eq:j_xi_theta} for both geometries and $J_{\xi_\phi}$~\eqref{eq:j_xi_phi} for the cube.

\subsection{Magnetic Current} 

\begin{figure}
\centering
\begin{subfigure}[b]{.49\textwidth}
\includegraphics[scale=.64,clip=true,trim=2.3in 0in 2.8in 0in]{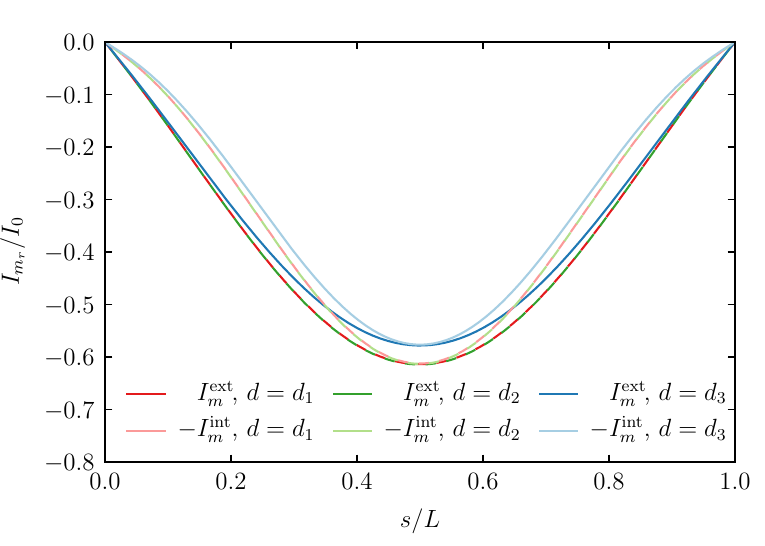}
\caption{$I_{m_r}$\vpad}
\end{subfigure}
\hspace{0.25em}
\begin{subfigure}[b]{.49\textwidth}
\includegraphics[scale=.64,clip=true,trim=2.3in 0in 2.8in 0in]{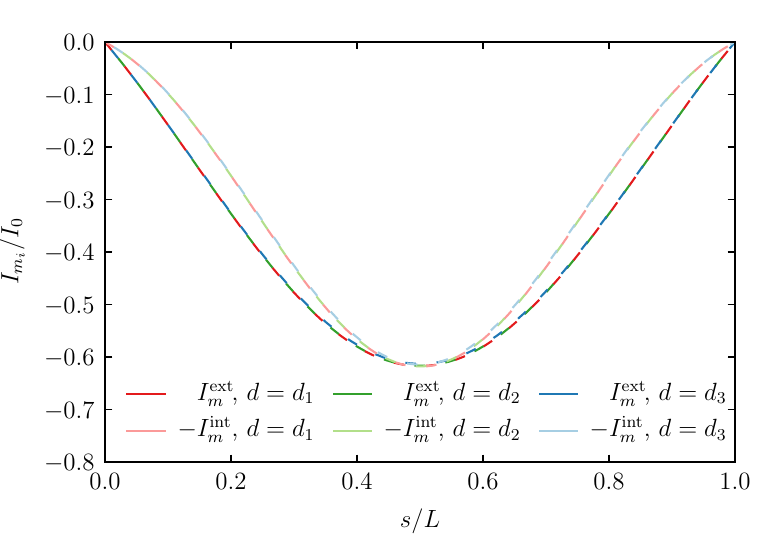}
\caption{$I_{m_i}$\vpad}
\end{subfigure}

\caption{$I_{m_{\modemaxp}}=I_{m_r}+jI_{m_i}$ for $d\in\{d_1,d_2,d_3\}$ and $\modemaxp=66$ in~\eqref{eq:Im_approx}.}
\vskip-\dp\strutbox
\label{fig:I_m}
\end{figure}

Rather than arbitrarily manufacturing $\mathbf{I}_\text{MS}$, we solve~\eqref{eq:no_mms_source} for  our choice of $\mathbf{J}_\text{MS}$.  $I_m(s)$ takes the form of~\eqref{eq:sine_series}, where, in~\eqref{eq:imqm} and~\eqref{eq:imqp}, $J_s(s)$ in~\eqref{eq:jsq} is
\begin{align}
J_s^\pm(s) =\pm J_0 f_{\xi_\theta}(\xi_w)g_{\eta_\theta}(\eta),
\label{eq:Js_choice}
\end{align}
where $s=\eta-\aslot$.  However, as stated in Section~\ref{sec:trunc}, it is necessary to approximate $I_m$ by truncating the infinite series.  We note that, for the geometry and choices of $J_s(s)$ we consider, $J_{s_\modeidxp}$~\eqref{eq:jsq} is zero for even values of $\modeidxp$; therefore, our approximation for $I_m$~\eqref{eq:sine_series} takes the form
\begin{align}
I_{m_{\modemaxp}}(s) = \sum_{\modeidxp=1}^\modemaxp I_{m_{\modeidxp'}}\sin \biggl(\frac{\modeidxp'\pi s}{\lslot}\biggr),
\label{eq:Im_approx}
\end{align}
where $\modeidxp'=2\modeidxp-1$.
Figure~\ref{fig:I_m} plots the real and imaginary components of $I_{m_{\modemaxp}}$ for each of the three depths, which are divided by $I_0=f_{\xi_\theta}(\xi_w)$~V for $\modemaxp=66$ in~\eqref{eq:Im_approx}.  

To determine how well~\eqref{eq:Im_approx} satisfies~\eqref{eq:slot_ext} and~\eqref{eq:slot_int}, we insert~\eqref{eq:Im_approx} into~\eqref{eq:slot_ext} and~\eqref{eq:slot_int}, which yields
\begin{align}
r_{\modemaxp}^-(s) &{}= {J_s^-(s)} + \frac{j\omega\epsilon}{4 w \bigl(k^2 -\beta_x^2\bigr)} \sum_{\modeidxp=1}^\modemaxp
\beta_{y_{\modeidxp'}}\sin \biggl(\frac{\modeidxp'\pi s}{\lslot}\biggr)
\bigl(
\bigl[I_{m_{\modeidxp'}}^+ - I_{m_{\modeidxp'}}^-\bigr] \tan(\beta_{y_{\modeidxp'}} d/2) + 
\bigl[I_{m_{\modeidxp'}}^+ + I_{m_{\modeidxp'}}^-\bigr] \cot(\beta_{y_{\modeidxp'}} d/2)
\bigr)
\nonumber\\
&{}= {J_s^-(s)} - J_{s_{\modemaxp}}^-(s)
\nonumber\\
&{}= -e_{J_{\modemaxp}}^-(s),
\label{eq:r_ext}
\\
r_{\modemaxp}^+(s) &{}= {J_s^+(s)} + \frac{j\omega\epsilon}{4 w \bigl(k^2 -\beta_x^2\bigr)} \sum_{\modeidxp=1}^\modemaxp
\beta_{y_{\modeidxp'}}\sin \biggl(\frac{\modeidxp'\pi s}{\lslot}\biggr)
\bigl(
\bigl[I_{m_{\modeidxp'}}^- - I_{m_{\modeidxp'}}^+\bigr] \tan(\beta_{y_{\modeidxp'}} d/2) + 
\bigl[I_{m_{\modeidxp'}}^+ + I_{m_{\modeidxp'}}^-\bigr] \cot(\beta_{y_{\modeidxp'}} d/2)
\bigl)
\nonumber\\
&{}= {J_s^+(s)} - J_{s_{\modemaxp}}^+(s)
\nonumber\\
&{}= -e_{J_{\modemaxp}}^+(s),
\label{eq:r_int}
\end{align}
where 
\begin{align*}
J_{s_{\modemaxp}}(s) = \sum_{\modeidxp=1}^\modemaxp J_{s_{\modeidxp'}}\sin \biggl(\frac{\modeidxp'\pi s}{\lslot}\biggr).
\end{align*}
For our manufactured $J_s(s)$~\eqref{eq:Js_choice}, $J_{s_{\modeidxp'}}$~\eqref{eq:jsq} is
\begin{align*}
J_{s_{\modeidxp'}}^- &{}= -J_0 f_{\xi_\theta}(\xi_w)\Biggl(\frac{\delta_{1{\modeidxp'}}}{2} - \frac{108\sqrt{3}{\modeidxp'}}
{\pi\bigl({81\modeidxp'}^4 -234{\modeidxp'}^2 +25\bigr)}\Biggr), 
\\
J_{s_{\modeidxp'}}^+ &{}= \phantom{-}J_0 f_{\xi_\theta}(\xi_w)\Biggl(-\frac{48\sqrt{2}(4{\modeidxp'}^3-17{\modeidxp'})}{\pi\bigl(64{\modeidxp'}^6-560{\modeidxp'}^4+1036{\modeidxp'}^2-225\bigr)}\Biggr).
\end{align*}
For sufficiently a large $\modeidxp'$,
\begin{align*}
\bigl|J_{s_{\modeidxp'}}\bigr|\le C_{J_{\modeidxp'}}{\modeidxp'}^{-3},
\end{align*}
such that $p=3$ in~\eqref{eq:jsq_limit}.

\begin{figure}
\centering
\begin{subfigure}[b]{.49\textwidth}
\includegraphics[scale=.64,clip=true,trim=2.3in 0in 2.8in 0in]{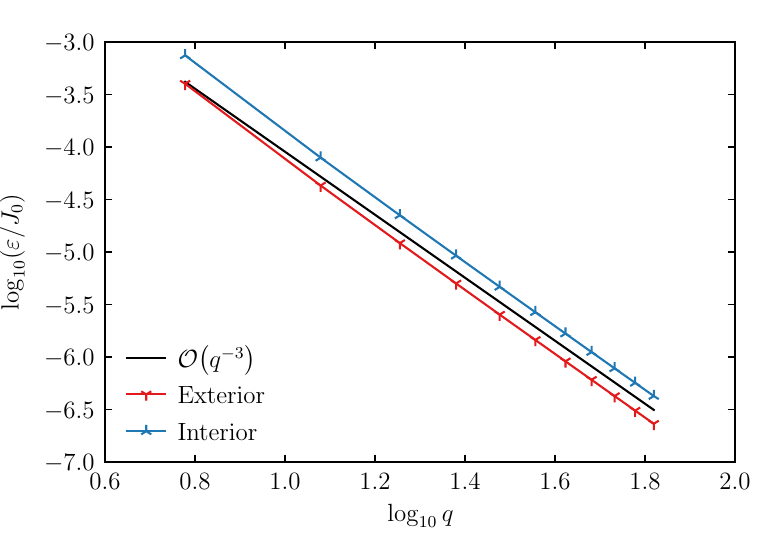}
\caption{$\varepsilon=|J_{s_{\modeidxp'}}|$\vpad}
\label{fig:J_coef_conv}
\end{subfigure}
\hspace{0.25em}
\begin{subfigure}[b]{.49\textwidth}
\includegraphics[scale=.64,clip=true,trim=2.3in 0in 2.8in 0in]{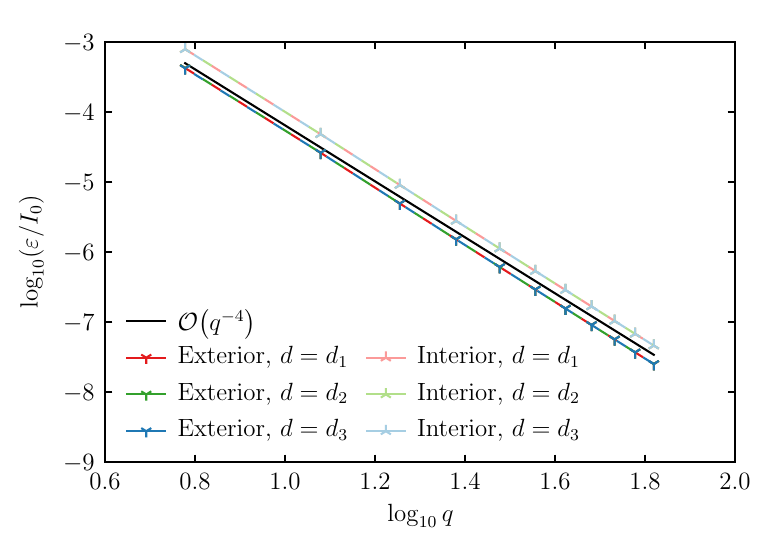}
\caption{$\varepsilon=|I_{m_{\modeidxp'}}|$\vpad}
\label{fig:Im_coef_conv}
\end{subfigure}
\caption{Sine series coefficient convergence.}
\vskip-\dp\strutbox
\label{fig:coef_conv}
\end{figure}

\begin{figure}
\centering
\includegraphics[scale=.64,clip=true,trim=2.3in 0in 2.8in 0in]{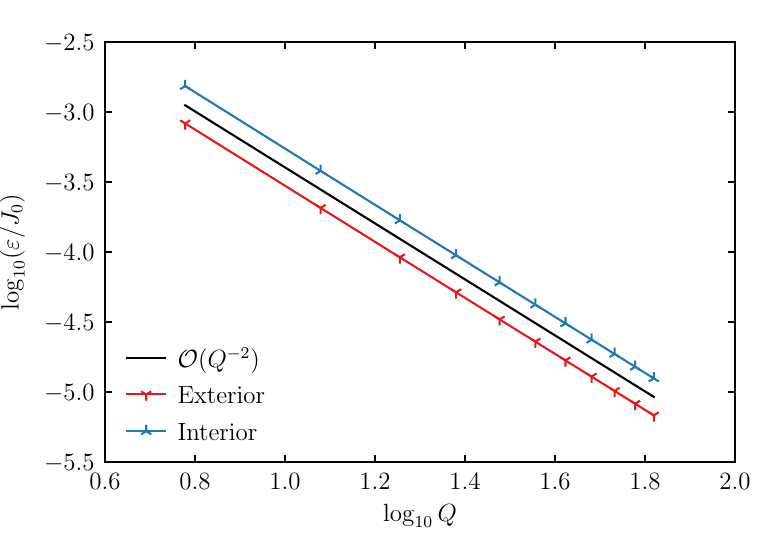}

\caption{Slot equation series truncation error: $\varepsilon=\|e_{J_{\modemaxp}}(s)\|_\infty$~\eqref{eq:eJQ_norm}.}
\vskip-\dp\strutbox
\label{fig:res_conv}
\end{figure}

Figure~\ref{fig:coef_conv} shows the convergence of the coefficients $J_{s_{\modeidxp'}}$ and $I_{m_{\modeidxp'}}$ with respect to ${\modeidxp}$, which are $\mathcal{O}({\modeidxp}^{-3})$ and $\mathcal{O}({\modeidxp}^{-4})$, respectively, as derived in Sections~\ref{sec:js} and~\ref{sec:im}.
Figure~\ref{fig:res_conv} shows the convergence of \eqref{eq:r_ext} and~\eqref{eq:r_int} with respect to $\modemaxp$ by measuring $\|e_{J_{\modemaxp}}(s)\|_\infty$~\eqref{eq:eJQ_norm}, which, as derived in Section~\ref{sec:js}, is $\mathcal{O}({\modemaxp}^{-2})$.

\subsection{Solution-Discretization Error} 
\label{sec:res_sde}

\begin{figure}[!t]
\centering
\begin{subfigure}[b]{.49\textwidth}
\includegraphics[scale=.64,clip=true,trim=2.3in 0in 2.8in 0in]{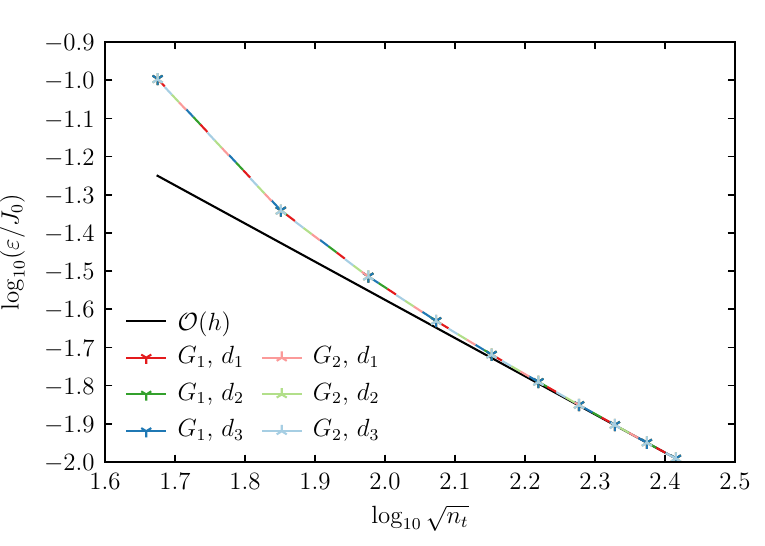}
\caption{$\modemaxp\sim\sqrt{n_t}$\vpad}
\label{fig:p1_errorJ_b4}
\end{subfigure}
\hspace{0.25em}
\begin{subfigure}[b]{.49\textwidth}
\includegraphics[scale=.64,clip=true,trim=2.3in 0in 2.8in 0in]{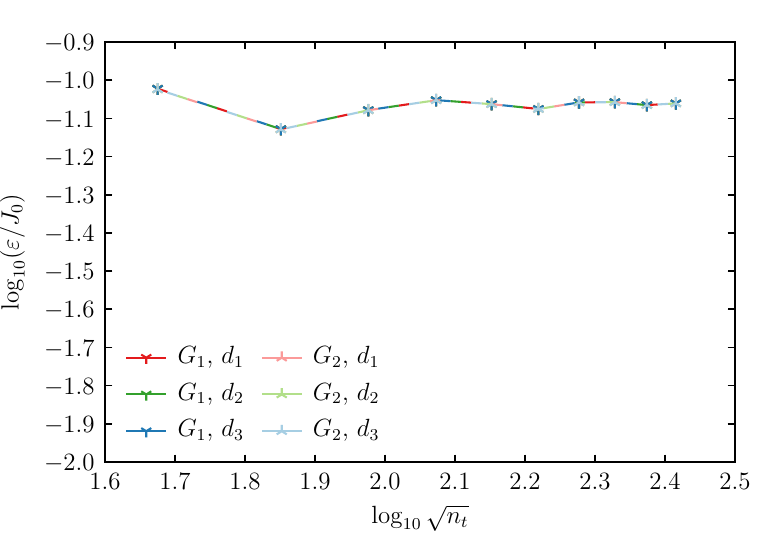}
\caption{$\modemaxp=1$\vpad}
\label{fig:p1_errorJ_b14}
\end{subfigure}
\caption{Solution-discretization error: $\varepsilon={\|\mathbf{e}_\mathbf{J}\|}_\infty$~\eqref{eq:solution_error_J} with discontinuity.}
\vskip-\dp\strutbox
\label{fig:p1_errorJ_b}
\end{figure}

\begin{figure}[!t]
\centering
\begin{subfigure}[b]{.49\textwidth}
\includegraphics[scale=.64,clip=true,trim=2.3in 0in 2.8in 0in]{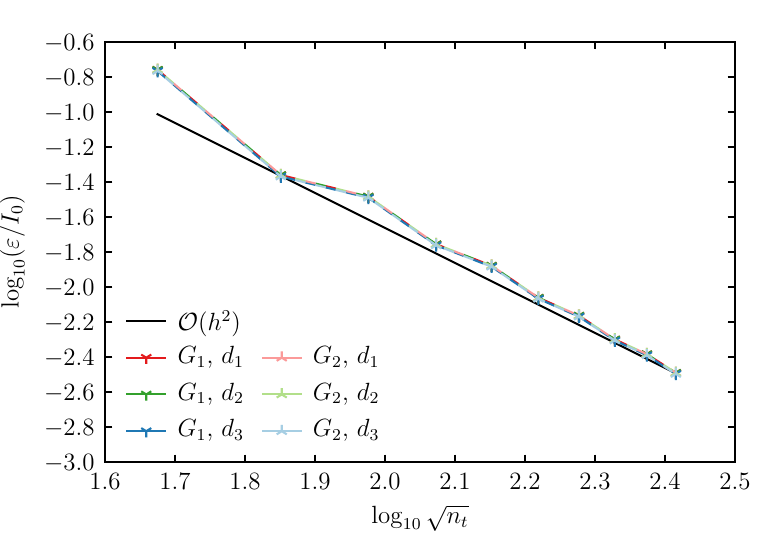}
\caption{$\modemaxp\sim\sqrt{n_t}$\vpad}
\label{fig:p1_errorI_b4}
\end{subfigure}
\hspace{0.25em}
\begin{subfigure}[b]{.49\textwidth}
\includegraphics[scale=.64,clip=true,trim=2.3in 0in 2.8in 0in]{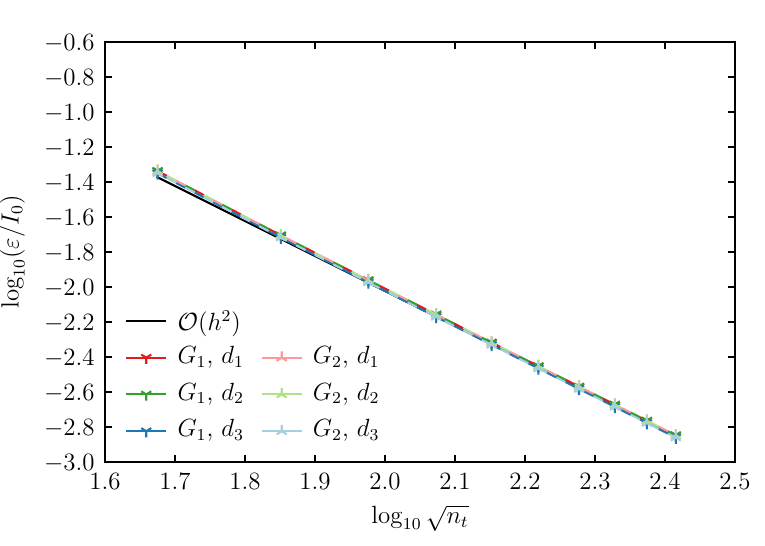}
\caption{$\modemaxp=1$\vpad}
\label{fig:p1_errorI_b14}
\end{subfigure}
\caption{Solution-discretization error: $\varepsilon={\|\mathbf{e}_\mathbf{I}\|}_\infty$~\eqref{eq:solution_error_I} with discontinuity.}
\vskip-\dp\strutbox
\label{fig:p1_errorI_b}
\end{figure}

\begin{figure}[!t]
\centering
\begin{subfigure}[b]{.49\textwidth}
\includegraphics[scale=.64,clip=true,trim=2.3in 0in 2.8in 0in]{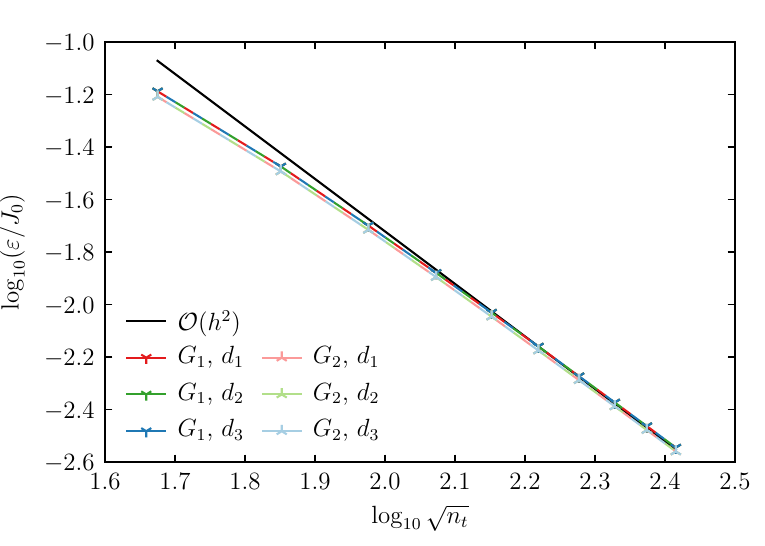}
\caption{$\modemaxp\sim\sqrt{n_t}$\vpad}
\label{fig:p1_errorJ_b0}
\end{subfigure}
\hspace{0.25em}
\begin{subfigure}[b]{.49\textwidth}
\includegraphics[scale=.64,clip=true,trim=2.3in 0in 2.8in 0in]{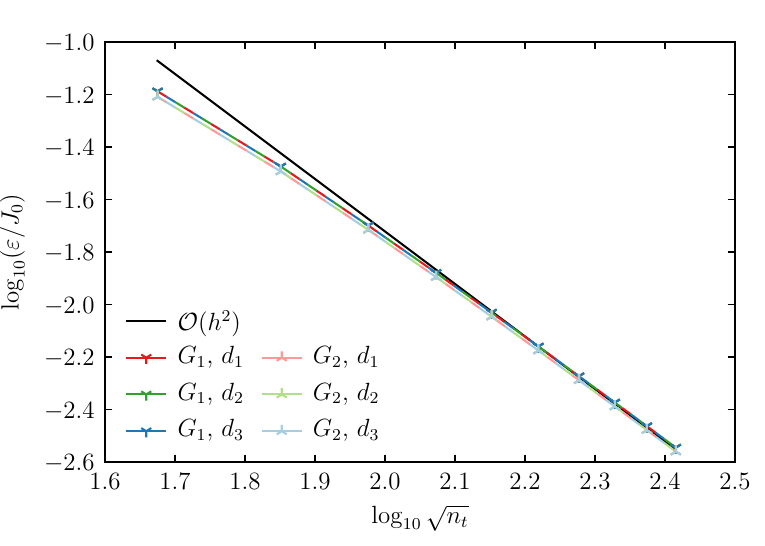}
\caption{$\modemaxp=1$\vpad}
\label{fig:p1_errorJ_b10}
\end{subfigure}
\caption{Solution-discretization error: $\varepsilon={\|\mathbf{e}_\mathbf{J}\|}_\infty$~\eqref{eq:solution_error_J} without discontinuity.}
\vskip-\dp\strutbox
\label{fig:p1_errorJ}
\end{figure}

\begin{figure}[!t]
\centering
\begin{subfigure}[b]{.49\textwidth}
\includegraphics[scale=.64,clip=true,trim=2.3in 0in 2.8in 0in]{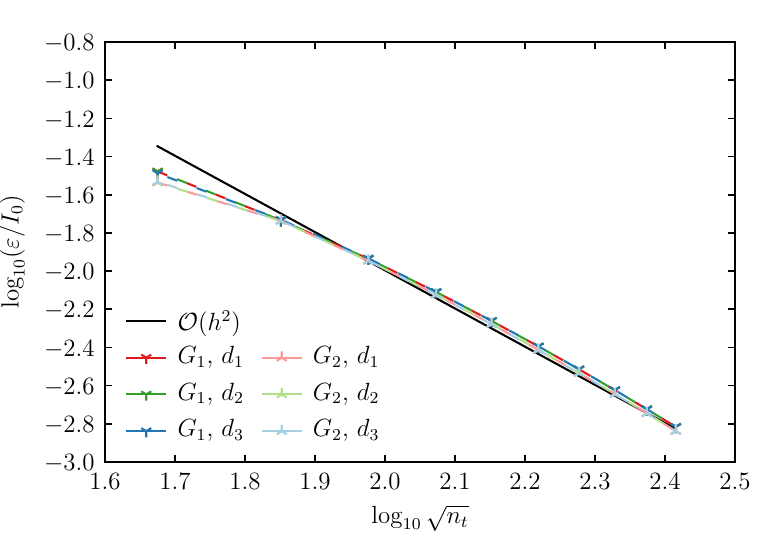}
\caption{$\modemaxp\sim\sqrt{n_t}$\vpad}
\label{fig:p1_errorI_b0}
\end{subfigure}
\hspace{0.25em}
\begin{subfigure}[b]{.49\textwidth}
\includegraphics[scale=.64,clip=true,trim=2.3in 0in 2.8in 0in]{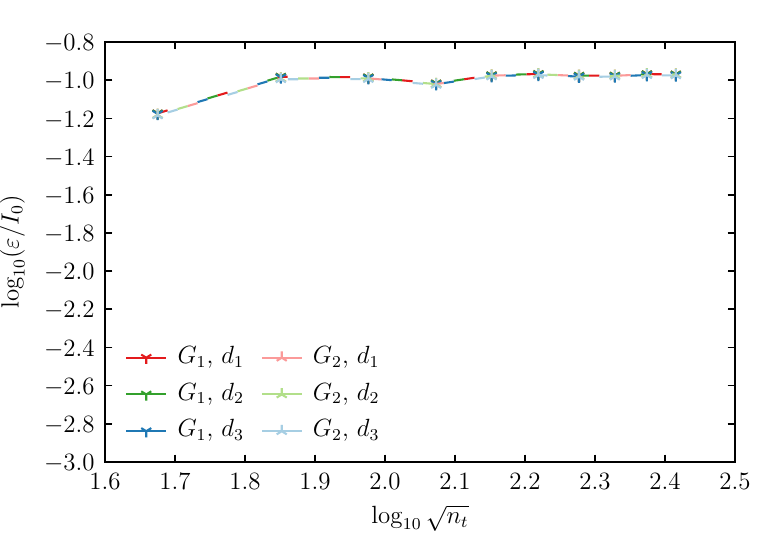}
\caption{$\modemaxp=1$\vpad}
\label{fig:p1_errorI_b10}
\end{subfigure}
\caption{Solution-discretization error: $\varepsilon={\|\mathbf{e}_\mathbf{I}\|}_\infty$~\eqref{eq:solution_error_I} without discontinuity.}
\vskip-\dp\strutbox
\label{fig:p1_errorI}
\end{figure}

Using the approaches in Section~\ref{sec:sde}, we exactly evaluate the integrals on both sides of~\eqref{eq:proj_disc_efie} and~\eqref{eq:proj_disc_slot} and measure the solution-discretization error.
The solution-discretization error arises from the basis-function approximation to the solution~\eqref{eq:sol_h}, as well as the truncation~\eqref{eq:Im_approx} of the sine series representation of $I_m$~\eqref{eq:sine_series}.  The convergence rate of the basis functions is expected to be $\mathcal{O}(h^2)$. From Section~\ref{sec:im}, the convergence rate of $\|e_{I_{\modemaxp}}(s)\|_\infty$~\eqref{eq:eIQ_norm} is expected to be $\mathcal{O}({\modemaxp}^{-3})$. Therefore, to measure the convergence of the solution-discretization error, it is sufficient to refine the series and mesh at the same rate ($\modemaxp\sim\sqrt{n_t}$).

Additionally, the linear system is solved using a matrix-ready generalized minimum residual (GMRES) method~\cite{saad_1986}.  Although Krylov-subspace methods are less frequently employed for dense matrices, GMRES is used here to allow parallelism to be extracted through distributed matrix--vector products.  200 basis vectors are used for all cases, which results in the ratio of the $L^2$-norm of the residual to the $L^2$-norm of the right-hand side being at most $10^{-13}$.  Since the condition numbers of the matrices are estimated to be $\mathcal{O}(10^7)$, this tolerance is expected to yield a relative error of no more than $\mathcal{O}(10^{-6})$ in the solution of the linear system.  Therefore, we expect negligible contamination from iteration error.

In this subsection, we show the $L^\infty$-norm of the discretization errors~\eqref{eq:solution_error_J} and~\eqref{eq:solution_error_I}: ${\|\mathbf{e}_\mathbf{J}\|}_\infty$ and ${\|\mathbf{e}_\mathbf{I}\|}_\infty$, which arise from only the solution-discretization error.  The error norms are shown for $G_\text{MS}\in\{G_1,\,G_2\}$~\eqref{eq:G_mms} and $d\in\{d_1,\,d_2,\,d_3\}$.

Figures~\ref{fig:p1_errorJ_b} and~\ref{fig:p1_errorI_b} show the convergence rates when the discontinuity described in Section~\ref{sec:sde} is present. For these assessments, we consider $\modemaxp=3\sqrt{n_t/140}$ in Figures~\ref{fig:p1_errorJ_b4} and~\ref{fig:p1_errorI_b4} and $\modemaxp=1$ in Figures~\ref{fig:p1_errorJ_b14} and~\ref{fig:p1_errorI_b14}.  
For the series in~\eqref{eq:ammsim} and~\eqref{eq:ammdissim}, we retain the first $2\modemaxp-1$ terms for $\modemaxp\sim\sqrt{n_t}$ and the first 150 terms for $\modemaxp=1$.  The latter is to ensure the matrix is sufficiently conditioned.
The convergence rate for ${\|\mathbf{e}_\mathbf{I}\|}_\infty$ in Figure~\ref{fig:p1_errorI_b4} is $\mathcal{O}(h^2)$, whereas the convergence rate for ${\|\mathbf{e}_\mathbf{J}\|}_\infty$  in Figure~\ref{fig:p1_errorJ_b4} is $\mathcal{O}(h)$, each as expected.  When $\modemaxp=1$, Figure~\ref{fig:p1_errorJ_b14} shows that ${\|\mathbf{e}_\mathbf{J}\|}_\infty$ does not decrease with mesh refinement.

Figures~\ref{fig:p1_errorJ} and~\ref{fig:p1_errorI} show the convergence rates when the discontinuity is removed.  In Figures~\ref{fig:p1_errorJ_b0} and~\ref{fig:p1_errorI_b0},  $\modemaxp=3\sqrt{n_t/140}$, and, in Figures~\ref{fig:p1_errorJ_b10} and~\ref{fig:p1_errorI_b10}, $\modemaxp=1$.  For $\modemaxp\sim\sqrt{n_t}$, the convergence rates for ${\|\mathbf{e}_\mathbf{J}\|}_\infty$ and ${\|\mathbf{e}_\mathbf{I}\|}_\infty$ are both $\mathcal{O}(h^2)$, as expected.  When $\modemaxp=1$, Figure~\ref{fig:p1_errorI_b10} shows that ${\|\mathbf{e}_\mathbf{I}\|}_\infty$ does not decrease with mesh refinement.  Figures~\ref{fig:p1_errorJ_b14} and~\ref{fig:p1_errorI_b10} underscore the importance of refining the series with the mesh.

\subsection{Numerical-Integration Error} 

Using the approaches of Section~\ref{sec:nie}, we measure the numerical-integration error, which arises from the use of generally approximate quadrature evaluations of the integrals~\eqref{eq:proj_disc_efie} and~\eqref{eq:proj_disc_slot}.  Depending on the element dimension, we consider either triangle or bar polynomial quadrature rules. 
As explained in Section~\ref{sec:im_int}, the integral of the truncation error associated with $I_{m_{\modemaxp}}$ is $\mathcal{O}({\modemaxp}^{-4})$.  Therefore, if $\modemaxp\sim 1/h$, where $1/h\sim \sqrt{n_t}$, as in Section~\ref{sec:res_sde}, the convergence rate will be limited to $\mathcal{O}(h^4)$.

\begin{figure}[!t]
\centering
\begin{subfigure}[b]{.49\textwidth}
\includegraphics[scale=.64,clip=true,trim=2.3in 0in 2.8in 0in]{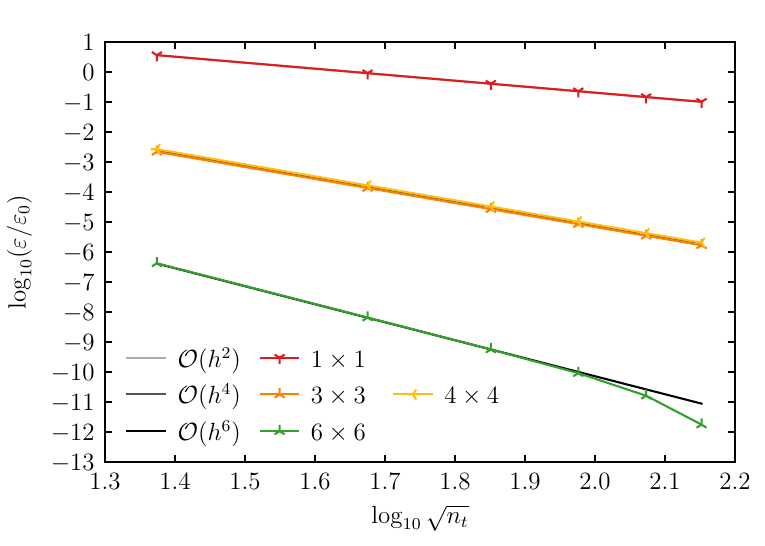}
\caption{$\modemaxp=1$\vpad}
\label{fig:part51}
\end{subfigure}
\hspace{0.25em}
\begin{subfigure}[b]{.49\textwidth}
\includegraphics[scale=.64,clip=true,trim=2.3in 0in 2.8in 0in]{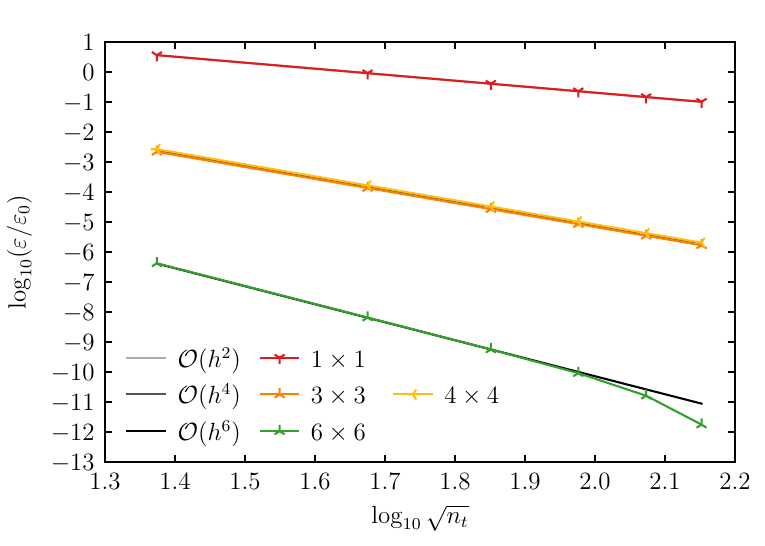}
\caption{${\modemaxp} \sim \sqrt{n_t}$ \vpad}
\label{fig:part52}
\end{subfigure}
\caption{Numerical-integration error: $\varepsilon=|e_a|$~\eqref{eq:a_error_cancel} for varying quadrature point amounts.}
\vskip-\dp\strutbox
\label{fig:part5}
\end{figure}

\begin{figure}[!t]
\centering
\begin{subfigure}[b]{.49\textwidth}
\includegraphics[scale=.64,clip=true,trim=2.3in 0in 2.8in 0in]{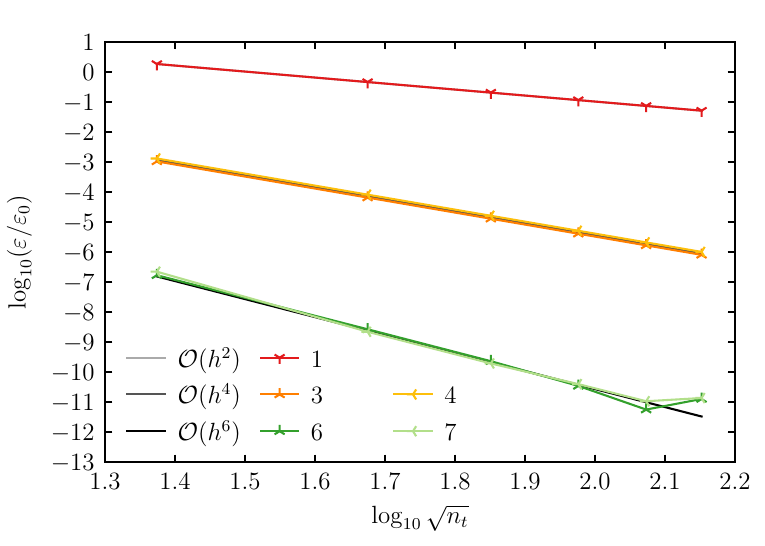}
\caption{$\modemaxp=1$\vpad}
\label{fig:part61}
\end{subfigure}
\hspace{0.25em}
\begin{subfigure}[b]{.49\textwidth}
\includegraphics[scale=.64,clip=true,trim=2.3in 0in 2.8in 0in]{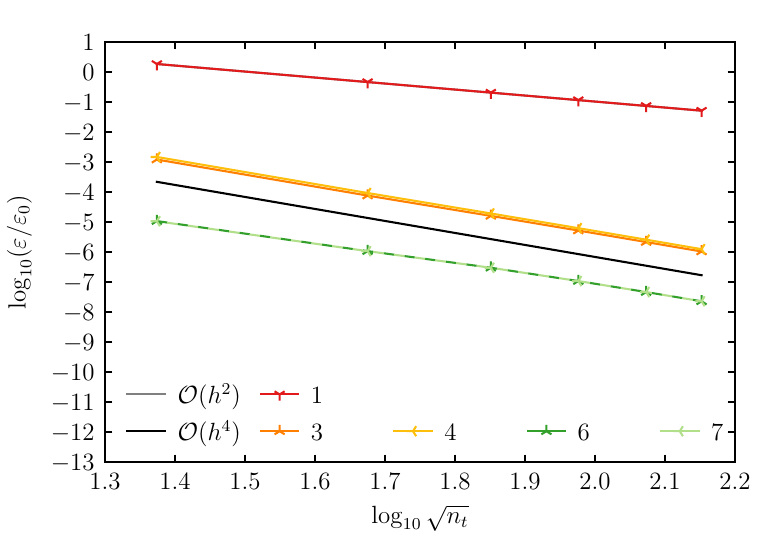}
\caption{${\modemaxp} \sim \sqrt{n_t}$ \vpad}
\label{fig:part62}
\end{subfigure}
\caption{Numerical-integration error: $\varepsilon=|e_b|$~\eqref{eq:b_error_cancel} for varying quadrature point amounts.}
\vskip-\dp\strutbox
\label{fig:part6}
\end{figure}

For $G_2$ and $d_1$, Figures~\ref{fig:part5} and~\ref{fig:part6} show the numerical-integration errors $e_a$~\eqref{eq:a_error_cancel} and $e_b$~\eqref{eq:b_error_cancel}.  
For each case, the amount of quadrature points is varied, with the legend entries taking the form $n_q^t\times n_q^s$, where $n_q^t$ is the amount of triangle quadrature points used to evaluate the test integrals and $n_q^s$ is the amount used for the source integrals.
We set the number of bar quadrature points to match the convergence rates of the triangle quadrature points.
We nondimensionalize $e_a$ and $e_b$ using the constant $\varepsilon_0=1$~A$\cdot$V.  
The entries in the left column of the legends are for reference convergence rates.  For a given legend row, we expect the simulation entries to convergence at the reference rate.

For these assessments, we consider $\modemaxp=1$ in Figures~\ref{fig:part51} and~\ref{fig:part61} and ${\modemaxp} = \sqrt{n_t/140}$ in Figures~\ref{fig:part52} and~\ref{fig:part62}. In Figures~\ref{fig:part51}, \ref{fig:part52} and~\ref{fig:part61}, the quadrature points converge at the expected rates, whereas, in Figure~\ref{fig:part62}, the convergence rates are limited to $\mathcal{O}(h^4)$, as expected, due to the integral of the truncation error associated with $I_{m_{\modemaxp}}$.  For the finest meshes and fastest convergence rate, the measurement of $e_a$ and $e_b$ is contaminated by the double-precision round-off error.

\section{Conclusions} 
\label{sec:conclusions}

In this paper, we presented methods for verifying the convergence rates due to the different interacting sources of numerical error when using the EFIE together with an arbitrary-depth slot model.
For the EFIE, we incorporated the manufactured surface current density through a manufactured incident field rather than through a dedicated source term.  Given this surface current, we derived a sine series representation for the associated magnetic current that satisfied the slot model equation exactly, obviating the need for a source term.

When measuring the solution-discretization error, we integrated exactly and avoided contamination from the sine series truncation and the iterative solver.  We refined the sine series truncation with the mesh and we kept the error due to the iterative solver sufficiently low.  
When measuring the numerical-integration error, we demonstrated the implications of the sine series truncation error on convergence.
For both approaches, we demonstrated expected convergence rates for several configurations.

\section*{Acknowledgments} 
\label{sec:acknowledgments}

This article has been authored by employees of National Technology \& Engineering Solutions of Sandia, LLC under Contract No.~DE-NA0003525 with the U.S.~Department of Energy (DOE). The employees own all right, title, and interest in and to the article and are solely responsible for its contents. The United States Government retains and the publisher, by accepting the article for publication, acknowledges that the United States Government retains a non-exclusive, paid-up, irrevocable, world-wide license to publish or reproduce the published form of this article or allow others to do so, for United States Government purposes. The DOE will provide public access to these results of federally sponsored research in accordance with the DOE Public Access Plan \url{https://www.energy.gov/downloads/doe-public-access-plan}.

\appendix
\renewcommand{\thesection}{\Alph{section}}

\section{The Arbitrary-Depth Slot Model}
\label{app:slot_model}

For a medium with finite electrical conductivity $\sigma$ and applying the Lorenz gauge condition and continuity equation, the electric field $\mathbf{E}$ and magnetic field $\mathbf{H}$ can be expressed in terms of the magnetic vector potential $\mathbf{A}$ and electric vector potential $\mathbf{F}$ in time-harmonic form as~\cite[Chap.~6]{balanis_2012}
\begin{align}
\mathbf{E} &{}= 
-\biggl(\frac{j}{\omega\mu\epsilon}\nabla(\nabla\cdot\mathbf{A}) + j\omega \mathbf{A} + \frac{1}{\epsilon}\nabla\times\mathbf{F}\biggr), 
\label{eq:E_pot}
\\
\mathbf{H} &{}= 
\frac{1}{\mu}\nabla\times\mathbf{A} -j\omega\mathbf{F} - \frac{j}{\omega\mu\epsilon}\nabla(\nabla\cdot\mathbf{F}),
\label{eq:H_pot}
\end{align}
where $\omega$ is the angular frequency, and $\mu$, $\sigma$, and $\epsilon$ are the permeability, conductivity, and potentially complex permittivity~\eqref{eq:comp_perm} of the medium. 

For a good electric conductor, the surface impedance boundary condition is~\cite[Chap.~14]{balanis_2012}
\begin{align}
\mathbf{E}-(\mathbf{E}\cdot\mathbf{n})\mathbf{n}=Z_s\mathbf{n}\times\mathbf{H},
\label{eq:sibc}
\end{align}
where $Z_s$ is the resistive surface impedance of the conductor, and $\mathbf{n}$ is the unit vector that is normal to the surface of the conductor and points away from the conductor.

The slot is modeled as a rectangular waveguide with an electrically small width~\cite{warne_2020}.  The waveguide supports transverse magnetic modes in the widthwise direction, such that $\mathbf{F}=\mathbf{0}$~\cite[Chap.~6]{balanis_2012}.  As a result, \eqref{eq:E_pot} and~\eqref{eq:H_pot} reduce to
\begin{align}
\mathbf{E} &{}= -\biggl(\frac{j}{\omega\mu\epsilon}\nabla(\nabla\cdot\mathbf{A}) + j\omega \mathbf{A}\biggr), \label{eq:E_A} \\
\mathbf{H} &{}= \frac{1}{\mu}\nabla\times\mathbf{A}. \label{eq:H_A}
\end{align}
In the absence of sources, $\mathbf{E}$ and $\mathbf{H}$ can be related through the Amp\`{e}re--Maxwell equation
\begin{align}
\nabla \times \mathbf{H} = j\omega\epsilon\mathbf{E}.
\label{eq:ampere}
\end{align}
Inserting~\eqref{eq:E_A} and~\eqref{eq:H_A} into~\eqref{eq:ampere} yields the Helmholtz equation
\begin{align}
\Delta \mathbf{A} + k^2 \mathbf{A} = \mathbf{0},
\label{eq:helmholtz}
\end{align}
where $k=\omega\sqrt{\mu\epsilon}$ is the wavenumber.


For notational convenience, we temporarily assume the rectangular waveguide is oriented such that the width, depth, and length are aligned with the $x$-, $y$-, and $z$-axes.
Because the widthwise dimension of the slot is assumed to be much smaller than the other two dimensions, the magnetic field is modeled as transverse magnetic to the widthwise direction.  Therefore, the magnetic vector potential takes the form $\mathbf{A}=A\mathbf{e}_x$~\cite{warne_2020}, such that \eqref{eq:E_A} becomes
\begin{align}
E_x = -\frac{j}{\omega\mu\epsilon}\biggl(\frac{\partial^2}{\partial x^2} + k^2\biggr) A, \qquad
E_y = -\frac{j}{\omega\mu\epsilon}\frac{\partial^2}{\partial x \partial y} A, \qquad
E_z = -\frac{j}{\omega\mu\epsilon}\frac{\partial^2}{\partial x \partial z} A;
\label{eq:E_comp}
\end{align}
\eqref{eq:H_A} becomes
\begin{align}
H_x = 0, \qquad
H_y =  \frac{1}{\mu} \frac{\partial}{\partial z}A, \qquad
H_z = -\frac{1}{\mu} \frac{\partial}{\partial y}A;
\label{eq:H_comp}
\end{align}
and \eqref{eq:helmholtz} becomes
\begin{align}
\Delta A + k^2 A = 0.
\label{eq:Ax_wave}
\end{align}
Equation~\eqref{eq:Ax_wave} is solved using separation of variables with $A$ taking the form
\begin{align}
A(x,y,z)= A_x(x) A_y(y) A_z(z), 
\label{eq:sep_var}
\end{align}
where
\begin{align}
A_\alpha(\alpha) = C_\alpha \cos(\beta_\alpha \alpha) + D_\alpha \sin(\beta_\alpha \alpha)
\label{eq:A_alpha}
\end{align}
for $\alpha\in\{x,\,y,\,z\}$.  $\beta_\alpha$ is the propagation constant in the $\alpha$-direction, and 
\begin{align}
k^2 = \beta_x^2 + \beta_y^2 + \beta_z^2.
\label{eq:k2}
\end{align}
With the electrically small width, $w\real(\beta_x)\ll 1$, such that, for $x\in[-w/2,\,w/2]$, $|\cos(\beta_x x)|\gg |\sin(\beta_x x)|$ in $A_x(x)$~\cite{warne_2020}, such that we can set $D_x=0$ in~\eqref{eq:A_alpha}.
The conducting surface of the scatterer yields the surface impedance boundary condition~\eqref{eq:sibc} for the medium.

\subsection{Widthwise Dependency}

For $x=\pm w/2$, $\mathbf{n}=\mp \mathbf{e}_x$, and, from~\eqref{eq:sibc}, $E_y=\pm Z_s H_z$ and $E_z=\mp Z_s H_y$, such that, from~\eqref{eq:E_comp} and~\eqref{eq:H_comp}, these boundary conditions are satisfied by
\begin{align}
\frac{d}{d x}A_x(\pm w/2) = \mp j Z_s \omega \epsilon A_x(\pm w/2).
\label{eq:beta_x_eq}
\end{align}
Equation~\eqref{eq:beta_x_eq} is satisfied by 
\begin{align}
\beta_x \tan(\beta_x w/2) = j Z_s \omega \epsilon.
\label{eq:beta_x_nonlinear}
\end{align}
Noting the electrically small width, $\beta_x \tan(\beta_x w/2) \approx w \beta_x^2/2 $, such that~\eqref{eq:beta_x_nonlinear} can be approximated by~\cite{warne_2020}
\begin{align}
\beta_x^2  \approx \frac{2 j Z_s \omega \epsilon}{w}.
\label{eq:beta_x_approx}
\end{align}
Ignoring the constant factor $C_x$ in~\eqref{eq:A_alpha}, we can approximate $A_x$ and $d^2 A_x/dx^2$ by~\cite{warne_2020}
\begin{align}
A_x(x) = \cos\beta_x x \approx 1, \qquad
\frac{d^2 A_x}{dx^2}   = -\beta_x^2 A_x(x) \approx -\beta_x^2.
\label{eq:Ax_approx}
\end{align}

\subsection{Lengthwise Dependency}
At $z=\{0,\,\lslot\}$, the electrically small width permits the conducting surface to be approximated as a perfect electric conductor, such that $H_z=0$~\cite{warne_2020}.  From~\eqref{eq:H_comp},
\begin{align}
A_z(0) = A_z(\lslot) = 0.
\label{eq:Az_bc}
\end{align} 
Equation~\eqref{eq:Az_bc} is satisfied by modes proportional to
\begin{align}
A_{z_\modeidx}(z) = \sin (\beta_{z_\modeidx}z),
\label{eq:Az}
\end{align}
where
\begin{align}
\beta_{z_\modeidx} =  \frac{\modeidx\pi}{\lslot},
\label{eq:beta_z}
\end{align}
for $\modeidx\in\mathbb{N}_1$.

\subsection{Depthwise Dependency}

With~\eqref{eq:Ax_approx} and~\eqref{eq:Az}, \eqref{eq:sep_var} becomes
\begin{align}
A(y,z) = \sum_{\modeidx=1}^\infty A_\modeidx(y,z),
\label{eq:A_series}
\end{align}
where
\begin{align}
A_\modeidx(y,z) = \bigl[C_{y_\modeidx} \cos(\beta_{y_\modeidx} y) + D_{y_\modeidx} \sin(\beta_{y_\modeidx} y)\bigr]\sin \biggl(\frac{\modeidx\pi z}{\lslot}\biggr).
\label{eq:A_n}
\end{align}

From~\cite{warne_1990}, at the inlet and outlet of the slot, the filament line-source magnetic current flowing along the wires in the lengthwise dimension is related to the voltage across the slot by
\begin{align}
\pm I_m^\pm(z)  = 2 V^\pm(z),
\label{eq:Im_V}
\end{align}
where the superscript ($-$) denotes the inlet ($y=-d/2$), and the superscript ($+$) denotes the outlet ($y=d/2$). 
The factor of 2 in~\eqref{eq:Im_V} is due to the convention used in~\cite{warne_2020} and~\cite{warne_1990}, where the magnetic current is doubled due to reflection in an infinite conducting plane (cf.~\cite[Chap.~7]{balanis_2012}).
%
%
The voltage across the slot is related to the electric field across the slot by~\cite{warne_2020}
\begin{align}
V^\pm(z) = w E_x(\pm d/2,z).
\label{eq:V_E}
\end{align}
From~\eqref{eq:A_series}--\eqref{eq:V_E} and \eqref{eq:E_comp},
\begin{align}
\pm I_m^\pm(z)/2 = \frac{-j w\bigl(k^2 -\beta_x^2\bigr)}{\omega\mu\epsilon} \sum_{\modeidx=1}^\infty \bigl[C_{y_\modeidx} \cos(\beta_{y_\modeidx} d/2) \pm D_{y_\modeidx} \sin(\beta_{y_\modeidx} d/2)\bigr]\sin \biggl(\frac{\modeidx\pi z}{\lslot}\biggr).
\label{eq:Im_1}
\end{align}
Multiplying~\eqref{eq:Im_1} by $\sin (\modeidxp\pi z / \lslot)$, integrating with respect to $z$, and noting that
\begin{align*}
\int_0^{\lslot}\sin \biggl(\frac{\modeidx\pi z}{\lslot}\biggr) \sin \biggl(\frac{\modeidxp\pi z}{\lslot}\biggr)dz = \frac{\lslot}{2} \delta_{\modeidx \modeidxp},
\end{align*}
where $\delta_{\modeidx \modeidxp}$ is the Kronecker delta, yields
\begin{align}
\pm\frac{1}{2}\int_0^{\lslot} I_m^\pm(z) \sin \biggl(\frac{\modeidx\pi z}{\lslot}\biggr) dz = 
\frac{-j w L\bigl(k^2 -\beta_x^2\bigr)}{2\omega\mu\epsilon} \bigl[C_{y_\modeidx} \cos(\beta_{y_\modeidx} d/2) \pm D_{y_\modeidx} \sin(\beta_{y_\modeidx} d/2)\bigr].
\label{eq:Im_2}
\end{align}
Adding the positive version of~\eqref{eq:Im_2} to the negative version of~\eqref{eq:Im_2} yields
\begin{align}
C_{y_\modeidx} = \frac{j\omega\mu\epsilon}{w L \bigl(k^2 -\beta_x^2\bigr) \cos(\beta_{y_\modeidx} d/2)}\int_0^\lslot \frac{1}{2}\bigl[I_m^+(z) - I_m^-(z)\bigr] \sin \biggl(\frac{\modeidx\pi z}{\lslot}\biggr) dz.
\label{eq:C_y_n}
\end{align}
Subtracting the negative version of~\eqref{eq:Im_2} from the positive version of~\eqref{eq:Im_2} yields
\begin{align}
D_{y_\modeidx} = \frac{j\omega\mu\epsilon}{w L \bigl(k^2 -\beta_x^2\bigr) \sin(\beta_{y_\modeidx} d/2)}\int_0^\lslot \frac{1}{2}\bigl[I_m^+(z) + I_m^-(z)\bigr] \sin \biggl(\frac{\modeidx\pi z}{\lslot}\biggr) dz.
\label{eq:D_y_n}
\end{align}
With the expressions for $\beta_x$~\eqref{eq:beta_x_approx} and $\beta_{z_\modeidx}$~\eqref{eq:beta_z}, $\beta_{y_\modeidx}$ can be obtained from \eqref{eq:k2}:
\begin{align}
\beta_{y_\modeidx}^2 = k^2 -\beta_x^2 - \beta_{z_\modeidx}^2.
\label{eq:beta_y}
\end{align}
From~\eqref{eq:C_y_n}, \eqref{eq:D_y_n}, and~\eqref{eq:beta_y}, $A$~\eqref{eq:A_series} and, consequently, $\mathbf{H}$ are expressed in terms of $I_m$.

\subsection{The Slot Equation}

To relate $I_m$ and $\mathbf{J}$, we consider the magnetic field at the openings of the slot, 
%
where the sum of the magnetic field due to $\mathbf{J}$ on the scatterer and the waveguide magnetic field is zero:
\begin{align}
\mathbf{J}^\pm\times\mathbf{n}^\pm - H_s(\pm d/2,s)\mathbf{s} = \mathbf{0}.
\label{eq:H_slot}
\end{align}
From~\eqref{eq:H_comp} and~\eqref{eq:A_series}, 
\begin{align*}
H_s(\pm d/2,s) ={}& -\frac{1}{\mu} \frac{\partial}{\partial y}A(\pm d/2,s) \nonumber
\\
={}& -\frac{j\omega\epsilon}{2 w L \bigl(k^2 -\beta_x^2\bigr)} \sum_{\modeidx=1}^\infty
\beta_{y_\modeidx}\int_{0}^{\lslot}\sin \biggl(\frac{\modeidx\pi s}{\lslot}\biggr)\sin \biggl(\frac{\modeidx\pi s'}{\lslot}\biggr)\times \nonumber
\\
& \bigl(
\pm \bigl[I_m^-(s') - I_m^+(s')\bigr] \tan(\beta_{y_\modeidx} d/2) + 
    \bigl[I_m^+(s') + I_m^-(s')\bigr] \cot(\beta_{y_\modeidx} d/2)
\bigr)ds'.
\end{align*}
Additionally, $I_m(0)=I_m(L)=0$.

To express the slot equation in its variational form, we project~\eqref{eq:H_slot} onto $\mathbb{V}^m$.  We seek the electric surface current density $\mathbf{J}\in\mathbb{V}$ and wire magnetic current $\mathbf{I}_m=I_m(s)\mathbf{s}\in\mathbb{V}^m$ that satisfy
\begin{align}
\int_0^\lslot \bar{\mathbf{v}}^m\cdot(\mathbf{J}^\pm\times\mathbf{n}^\pm) ds 
{}+{}
\frac{j\omega\epsilon}{2 w L \bigl(k^2 -\beta_x^2\bigr)}
\sum_{\modeidx=1}^\infty
\beta_{y_\modeidx}\int_0^\lslot \bigl(\bar{\mathbf{v}}^m(s)\cdot\mathbf{s}\bigr) \sin \biggl(\frac{\modeidx\pi s}{\lslot}\biggr)ds\int_{0}^{\lslot}\sin \biggl(\frac{\modeidx\pi s'}{\lslot}\biggr)\times \nonumber
\\
\bigl(
\pm \bigl[I_m^-(s') - I_m^+(s')\bigr] \tan(\beta_{y_\modeidx} d/2) + 
    \bigl[I_m^+(s') + I_m^-(s')\bigr] \cot(\beta_{y_\modeidx} d/2)
\bigr)ds'
{}= 0
\label{eq:slot_variational}
\end{align} 
for every $\mathbf{v}^m\in\mathbb{V}^m$.

\addcontentsline{toc}{section}{\refname}
\bibliographystyle{elsarticle-num}
\bibliography{../quadrature_manuscript/quadrature.bib}

\end{document}